\documentclass[aps,pre,citeautoscript,singlecolumn,showkeys, floatfix, superscriptaddress, notitlepage]{revtex4-1}
\pdfoutput=1
\usepackage{amsmath}
\usepackage{amssymb}
\usepackage{hyperref}
\usepackage{mathtools}
\usepackage{empheq}
\usepackage{graphicx}
\usepackage{float}
\usepackage{enumitem}
\usepackage{layout}
\usepackage{subfig}
\usepackage{natbib}
\usepackage{epstopdf}
\usepackage{amsthm}
\usepackage{xcolor}
\usepackage{textcase}
\usepackage{csquotes}
\usepackage[normalem]{ulem}
\usepackage{epigraph}
\usepackage{subfiles}

\let\e=\epsilon

\def\to{\rightarrow}

\newcommand{\beq}{\begin{equation}} \newcommand{\eeq}{\end{equation}}

\hypersetup{
    colorlinks=true,
    linkcolor=blue,
    filecolor=magenta,      
    urlcolor=cyan,
}

\begin{document}


\title{Good speciation and endogenous business cycles in a constraint satisfaction macroeconomic model}

\author{\underline{Dhruv Sharma}}
\thanks{Corresponding author. Email: dhruv.sharma@polytechnique.org}
\affiliation{Laboratoire de Physique de l'Ecole Normale Sup\'erieure, ENS, Universit\'e PSL, CNRS, Sorbonne Universit\'e, Universit\'e de Paris, Paris, France}
\affiliation{Chair of Econophysics \& Complex Systems, Ecole polytechnique, 91128 Palaiseau, France}

\author{Jean-Philippe Bouchaud}
\affiliation{CFM, 23 rue de l'Universit\'e, 75007 Paris, France
}
\affiliation{Chair of Econophysics \& Complex Systems, Ecole polytechnique, 91128 Palaiseau, France}

\author{Marco Tarzia}
\affiliation{
	LPTMC, CNRS-UMR 7600, Sorbonne Universit\'e, 4 Pl. Jussieu, F-75005, Paris, France
}
\affiliation{Institut  Universitaire  de  France,  1  rue  Descartes,  75231  Paris  Cedex  05,  France}

\author{Francesco Zamponi}
\affiliation{Laboratoire de Physique de l'Ecole Normale Sup\'erieure, ENS, Universit\'e PSL, CNRS, Sorbonne Universit\'e, Universit\'e de Paris, Paris, France}

\begin{abstract}
We introduce a prototype agent-based model of the macroeconomy, 
with budgetary constraints at its core. The model is related to a class of constraint satisfaction problems (CSPs), which has been thoroughly investigated in computer science. The CSP paradigm allows us to propose an alternative price-setting mechanism: given agents' preferences and budgets, what set of prices satisfies the maximum number of agents? Such an approach permits the coupling of production and output within the economy to the allowed level of debt in a simplified framework. Within our model, we identify three different regimes upon varying the amount of debt that each agent can accumulate before defaulting. In presence of a very loose constraint on debt, endogenous crises leading to waves of synchronized bankruptcies are present.
In the opposite regime of very tight debt constraining,
the bankruptcy rate is extremely high and the economy remains structure-less. 
In an intermediate regime, the economy is stable with very low bankruptcy rate and no aggregate-level crises. 
This third regime displays a rich phenomenology:
the system spontaneously and dynamically self-organizes in a set of cheap and expensive goods (i.e. some kind of ``speciation''), with switches triggered by random fluctuations and feedback loops. Our analysis confirms the central role that debt levels play in the stability of the economy. More generally, our model shows that constraints at the individual scale can generate highly complex patterns at the aggregate level.  
\end{abstract}
\keywords{endogenous cycles; constraint-satisfaction problems; agent-based models; speciation; macroeconomics; debt dynamics;} 
\maketitle



\section{Introduction}

Neo-classical economics, based on the idea of fully rational representative agents who optimize their intertemporal utility function, held a hegemonic position in academic and institutional circles for years until its serious shortcomings were brutally brought to the fore during the crisis of 2008. Calls for ``rebuilding macroeconomics'' can be heard across the board (central banks, IMF, OECD, and of course university economics departments) ~\cite{Vines2017, Blanchard2018}. 

Once one abandons the straitjacket of rational expectations, so many possibilities open up that the modelling endeavour seems hopeless to many -- as Sims and Sargent noted, abandoning rational expectations leads a modeler into the ``wilderness of bounded rationality''~\cite{Sims1980, Sargent1993}. There is another way to look at how research should be conducted, though, which turns the ``wilderness'' predicament on its head. Unshackled by the immediate necessity of logical consistency and direct empirical relevance, modelers can explore new ideas that suggest explanations for real world {\it phenomena} that are unfathomable within the context of classical approaches. One of the most baffling such so-called ``anomalies'', is the ``excess volatility'' puzzle (also called ``small shocks, large business cycle'' puzzle)~\cite{Shiller1981, Long1983, Bernanke1996, Cochrane1994}: fluctuations of both economic activity and financial prices seem way too large to be accountable by ``dynamic stochastic general equilibrium'' models (DSGE). 

The understanding and modeling of the phenomena that can emerge beyond general equilibrium should benefit from the remarkable progress made in other fields of ``complex systems'' research (physics, computer science, ecology, etc.). One came to realize that non-linear, interacting units can lead to a variety of interesting phenomena, like phase transitions~\cite{Ma2018, Goldenfeld1992, Challet2005, Gualdi2015a}, complex sustained endogenous dynamics and synchronisation effects~\cite{Watts1998, Gualdi2015b}, history dependence~\cite{Sethna2001}, intrinsic fragility and sudden breakdown~\cite{Bak2013, Bouchaud2013}, to name a few. It is difficult not to think that these effects, that all show up in socio-economic systems, are governed by some of the general mechanisms present in complex systems~\cite{Epstein1999, Goldstone2005} (for recent reviews see e.g.~\cite{Castellano2009, Bouchaud2013}). This view has of course been held by many authors since the famous 1987 Santa Fe conference ``Economics as a complex evolving system''~\cite{Anderson1988}. But for some reason, these ideas have not really made it yet to the mainstream. Still, the study of ``Agent Based Models'' (ABM) to understand economic systems has been booming in the last ten years, with many complex systems ideas finding a natural framework to express themselves~\cite{Gualdi2015a, Turrell2016, Braun-Munzinger2018,Baptista2016, Lamperti2018a, Haldane2018}.

Among the most important, overarching models of complex systems are the so-called ``Constraint Satisfaction Problems'' (CSP). In a nutshell, these problems contain a certain number of variables (for example binary variables) that must satisfy certain linear or non-linear constraints, like equalities or inequalities. $K$-SAT problems, for example, require $N$ binary variables to simultaneously satisfy $M$ logical clauses, each clause containing $K$ variables~\cite{handbook}. Typically, such problems have solutions when the number of clauses $M$ is sufficiently small compared to the number of variables (SAT phase), but run into contradictions when $M$ is too large (UNSAT phase). Interestingly, the transition between these two regimes becomes sharp in the limit $N,M \to \infty$ with $\alpha = M/N$ fixed~\cite{Mezard2002c}. Furthermore, the landscape of solutions becomes extremely complex in the SAT region close to the phase transition: solutions become rare, and are clustered in disconnected ``regions'' of the variables' space~\cite{Altarelli2008a, Antenucci2018}. Some of these clusters may suddenly disappear as $M$ increases, forcing the system to re-adapt in a completely different configuration -- a crisis of sorts. 

Since economic problems very often involve several types of constraints, such as budget constraints, leverage constraints or non-substitutability effects, we believe that the CSP paradigm would lead to important modelling insights in economic situations as well, where constraints by themselves can generate non-trivial aggregate effects. 

The aim of the present paper is to explore this idea in a simple setting: a market of goods with producers and consumers. Agents in the economy have to satisfy budget and/or production constraints, and they adjust their strategies to optimize their profit while satisfying these constraints. Crucially, within the CSP paradigm, we optimize a ``global'' cost function that minimizes the total number of unsatisfied constraints, and not satisfy each agents' constraints individually. In other words, prices are set in such a way that the least number of agents' constraints is unsatisfied.
Formally, our model is identical to the so-called ``perceptron model'', another classical and well studied CSP (see \ref{subsec:perceptron} below), which can show exponentially many equilibria in certain regions of parameter space~\cite{Franz2017a}. However, in this work we consider a regime in which there is always a single equilibrium to the perceptron problem. Nonetheless, our CSP-inspired macroeconomic ABM displays many interesting features, such as phase transitions, speciation, endogeneous business cycles, etc. 
We also believe that this model opens the way to investigate more complex variants, which could in principle display even more interesting phenomena such as an exponential proliferation of equilibria and, correspondingly, non-ergodic behaviour. 


\section{\label{sec:model}The Economy as a constraint satisfaction problem}

In our model, we consider $M$ agents and $N$ products with prices $p_{1} \dotsc p_{N}$. Each agent, labeled by $\mu = 1 \dotsc M$ wants to buy or sell a certain quantity $\xi_{\mu}^{i}$ of the product $i$ (positive for selling the product and negative for buying it).
We normalize the prices such that at all times $\frac{1}{N} \sum_{i} p_{i} =1$ and that all prices are bounded from below $p_{i} \geq x_{m}$, where $x_{m}$ is a non negative real number, possibly zero.
The evolution of prices is given by a ``market'' that attempts to enforce simple budgetary constraints for all agents, while the evolution of preferences is decided by the agents according to a simple behavioral rule, as we now describe.

\subsection{\label{subsec:prices} Budget constraint and formation of prices}

 The quantity $\pi_{\mu} = \sum_{i} \xi_{\mu}^{i} p_{i}$ is the total money {that the agent $\mu$ is willing to spend (or earn)} 
 in the market in a given round. This quantity is subject to a budget constraint, namely that $\pi_{\mu} \geq \sigma$. Here, $\sigma < 0$ if the agent is allowed to borrow to cover losses and $\sigma >0$ if the agent is required to make a profit. 
The products prices $p_{i}$ thus depend dynamically on agents' preferences. Given the matrix of preferences $\vec{\xi}_{\mu}$, we assume that prices are determined by the market, in such a way that the least number of agents have unsatisfied budgetary constraints. We define a variable called the ``gap'' as follows:
\begin{align}\label{define_h_mu}
h_{\mu}(\vec{p}):= 
\vec \xi_{\mu} \cdot \vec p 
-  \sigma > 0 \qquad \forall \mu \in \left \{ 1 \ldots M \right\} \, .
\end{align}
The gap variable hence encodes the distance from the configuration where the budgetary constraint for agent $\mu$ is on the verge of being unsatisfied ($h_{\mu} =0$). The price vector $\vec{p}$ is then determined by minimizing the number of agents whose constraints are unsatisfied. This can be achieved by minimizing the following cost function:
\begin{align}
\mathrm{H}(\vec{p}) = \frac{1}{2} \sum_{\mu=1}^{M} h_{\mu}^{2} \, \Theta(-h_{\mu}) \, ,
\label{eq:constraint}
\end{align}
under the constraints that $\frac{1}{N} \sum_{i} p_{i} =1$ and $p_{i} \geq x_{m}$, where $x_{m}$ is a small positive number thus ensuring that all prices are positive.  The Heaviside $\Theta$ function is equal to 1 when its argument is positive, and zero otherwise. 
This function (or Hamiltonian) takes its minimal value (i.e. zero) when all agents are satisfied and is positive if at least one agent is unsatisfied. In what follows, we use the formulation ``unsatisfied agent'' to mean that the agent's budgetary constraint is presently unsatisfied.

Note that the constraint on prices $\frac{1}{N} \sum_{i} p_{i} =1$ sets the price units. Since it is a linear constraint, the above optimisation problem is convex and the solution space remains connected. This is no longer the case for other types of constraints, for example $\frac{1}{N} \sum_{i} (p_{i} - p_i^0)^2 =1$ where $p_i^0$ are some reference prices (as considered in \cite{Sharma2019}, see also section \ref{subsec:perceptron}) or $\frac{1}{N} \sum_{i} \vert p_{i} - p_i^0 \vert \leq \Gamma$ (corresponding to repricing costs). In such cases, one may anticipate an even richer phenomenology, but we will not consider them further in the present study.

\subsection{\label{subsec:preferences}Preferences update: supply and demand}

Agents' preferences $\vec{\xi}_{\mu}$ are allowed to evolve in reaction to supply/demand imbalances and prices, according to a simple behavioral rule similar to those used in the ``MarkI'' and ``Mark0'' models~\cite{Gaffeo2008,Gualdi2015a}. Contrary to the infinite horizon, profit maximizing framework which is \emph{de rigueur} in standard microeconomic models, we posit reasonable, heuristic rules to model the behavior of agents. These rules take the following form: 

\begin{itemize}
	\item \underline{Supply side}. If the agent is a supplier for product $i$ ($\xi^i_{\mu} >0$), then it adapts as a function of the mismatch between the supply $S_{i}$ and demand $D_{i}$  of product $i$, defined as:
	\begin{align}
	\label{def:supply} 
	S_{i} = \sum_{\mu} \xi_{\mu}^{i} \, \Theta(\xi_{\mu}^{i}) \, , \qquad D_{i} = \sum_{\mu} \xi_{\mu}^{i} \, \Theta(-\xi_{\mu}^{i}) \, .
	\end{align}
	The supplier updates their preference as: 
	\begin{align} 
	\label{def:update1} 
	\begin{split}
 	S_{i} > D_{i} &\implies \xi_{\mu}^{i}(t+1) = \xi_{\mu}^{i}(t)(1-\epsilon_{D} {\tt u}) \, ,\\
 	S_{i} <  D_{i} &\implies \xi_{\mu}^{i}(t+1) = \xi_{\mu}^{i}(t)(1+\epsilon_{D} {\tt u}) \, ,
	\end{split}
 	\end{align}
 	where ${\tt u}$ is a random number sampled independently from the uniform distribution on \([0,1]\), and \(\epsilon_{D}\) is the speed of adjustment to supply-demand pressure.
\item \underline{Demand side}. If the agent is a buyer for product $i$ ($\xi^i_{\mu} < 0$), it adapts its preferences looking at the relative price level for the product as follows:
	\begin{align}
	\label{def:update2} 
	\begin{split}
 	p_{i} > 1  &\implies \xi_{\mu}^{i}(t+1) = \xi_{\mu}^{i}(t)(1-\epsilon_{P} {\tt u}) \, ,\\
  	p_{i} < 1  &\implies \xi_{\mu}^{i}(t+1) = \xi_{\mu}^{i}(t)(1+\epsilon_{P} {\tt u}) \,  ,
	\end{split}
	\end{align}
	where once again \({\tt u}\) is sampled independently from the uniform distribution and \(\epsilon_{P}\) denotes the speed of adjustment to price pressure.

\end{itemize}

\subsection{\label{subsec:transactions}Transactions, production costs and redistribution}

Agents then perform transactions, which determine the actual quantities of products sold and bought. The update rules depend on the ratio of supply and demand for product $i$, noted $\zeta_i = \frac{S_{i}}{D_{i}}$. If \(\zeta_{i} >1\), then the suppliers in the market are unable to dispose off all their inventory and sell only a fraction of it. Conversely if \(\zeta_{i}  <1\), then the buyers are unable to satisfy their demands for product \(i\).
These transactions are encoded in the matrix of realized supply/demand \(\bar{\xi}_{\mu}^{i}\), which takes the following form:
\begin{align}
\begin{split}
\zeta_{i} > 1, \xi_{\mu}^{i} >0 &\implies \bar{\xi}_{\mu}^{i} = \frac{\xi_{\mu}^{i}}{\zeta_{i}} \, ,\\
\zeta_{i} < 1, \xi_{\mu}^{i} <0 &\implies \bar{\xi}_{\mu}^{i} = \zeta_{i} \xi_{\mu}^{i} \, .
\end{split}
\end{align}
The matrix \(\bar{\xi}_{\mu}^{i}\) hence enters the computation of the true money exchanged by the agents. 

We posit that each product has a production cost $\gamma_{i}$ associated with it, paid by all the suppliers of product \(i\) and sampled from a uniform distribution \([0, \gamma]\). 
The production cost of all goods in the economy is redistributed to the agents as wages. We define the total production cost as 
$W(t) = \sum_{\mu, i} \gamma_{i} \xi^{\mu}_{i}(t) \, \Theta(\xi_{\mu}^{i}(t))$. This is then uniformly distributed to all agents with each agent getting $w = \frac{W}{M}$ back as ``wage''.

The ``full'' profit 
(or money exchanged) at time $t$ by agent $\mu$ thus reads:
\begin{equation}
\label{eq:money_exchanged2}
\bar\pi_{\mu}(t) = \sum_{i} \bar{\xi}_{\mu}^{i} p_i(t) - \sum_{i} \gamma_{i} \xi^{\mu}_{i} \, \Theta(\xi_{\mu}^{i}) + w(t) \, ,
\end{equation}
where the first term corresponds to transactions, the second to production costs, and the third to the wage.
Note that because of the ``market clearing'' condition, $\sum_\mu \bar{\xi}_{\mu}^{i}=0$, we have
$\sum_\mu \bar\pi_{\mu}(t)=0$, so that the total amount of circulating money is conserved.
Note also that the production cost and wage terms, on average, compensate each other. This is the reason why we do not take them into account in the price formation process described in section~\ref{subsec:prices}.

\subsection{\label{subsec:removal}Removal and Replacement of agents}

At the end of these steps, it is possible that there remain some agents whose budget constraints are unsatisfied. The economic interpretation of an unsatisfied agent is straightforward: the agent cannot participate in the economy and should in principle go bankrupt and be removed. However, instead of looking at the instantaneous value of the constraint (which could be sensitive to random local fluctuations), we impose that the budget be satisfied on average over some time window. 

More precisely, we take an exponentially moving average of the money that agents exchange and compare it to their budget. We define $\pi_{\mu}^{\text{ema}}(t)$ which is the averaged money exchanged over multiple time steps (over a duration of the order of $\omega^{-1}$): 
\begin{equation}
\label{money_ema} 
\pi_{\mu}^{\text{ema}}(t) = \omega \bar{\pi}_{\mu}(t) + (1-\omega) \pi_{\mu}^{\text{ema}}(t-1) \, .
\end{equation}
The agent $\mu$'s business is deemed unsustainable if $\pi_{\mu}^{\text{ema}}(t)<\sigma$, and the agent is then replaced by a new agent. The new agents' preferences $\vec{\xi}^{\mu}$ are sampled independently from the Normal distribution $\mathcal{N}(0,1)$.
We also initialize $\pi_{\mu}^{\textrm{ema}}$ for the new agent with the average $\bar{\pi}_{\mu}$ of the remaining surviving agents. 
Note that when $\omega=1$, agents violating their budget constraint are immediately removed.

\subsection{Summary of the parameters}

With the dynamical rules above, our model thus has the following parameters:
\begin{enumerate}
\item \(\alpha = \frac{M}{N}\) : the ratio of the number of agents to the number of products
\item \(\sigma\): the budgetary constraint
\item \(\epsilon_{D}\), \(\epsilon_{P}\): susceptibilities of the agents to demand and price pressures
\item \(\gamma\): parameter of the uniform distribution which fixes the production cost of the products
\item \(\omega^{-1}\): timescale {over which the agents' average profit is computed.} 

\end{enumerate}
The model is completely specified once these parameters are provided along with the initial distributions for the price vector $\vec{p}$ and agents' preferences $\vec{\xi}_{\mu}$. 


\subsection{\label{subsec:perceptron}Relation to the perceptron model}

The Hamiltonian in Eq.~\eqref{eq:constraint} is formally identical to that of the perceptron model, as formulated e.g. in Ref.~\cite{Franz2017a}. 
The perceptron is a classic machine learning problem: 
it was one of the first formal models proposed to understand how neurons function, and has been the object of study for many decades. Starting with the pioneering work of Rosenblatt~\cite{Rosenblatt1958}, it has continued to be both a paradigm and a cornerstone of machine learning. Statistical physicists became interested in the perceptron problem and significant progress was made in the late 1980’s in a series of papers~\cite{Gardner1988, Gardner1988a, Krauth1988, Krauth1989, Brunel1992}. Recently, interest in the perceptron surged again within the physics community after Franz and Parisi proposed it as a toy model to understand the jamming of hard spheres (see~\cite{Franz2017a} for a detailed discussion). In the simplest setting, the disorder (agents' preferences $\vec{\xi}^{\mu}$ here) is considered as quenched, while the variables are encoded in the vector $\vec{p}$ living on the $N$-dimensional hypersphere. The models are then studied in the thermodynamic limit with a Hamiltonian equivalent to Eq.~\eqref{eq:constraint}. With the imposition of the spherical constraint on $\vec{p}$, the perceptron problem becomes non-convex for negative $\sigma$ and a rich phase diagram is obtained~\cite{Franz2017a}. In this work however, agents' preferences evolve as price changes and hence we do not work with quenched disorder, but time dependent (annealed) disorder. Furthermore, we impose a linear constraint and a positivity constraint on the price vector, which makes the optimisation problem convex. As mentioned above, non linear generalisations would be interesting to study further, along the lines of \cite{Sharma2019,Marsili2014}.

\section{Results}

In this section, we present the results obtained through numerically simulating the dynamics of the model. In what follows and unless otherwise noted, we set $N=100$ products, $\alpha=10$ (i.e. $M=1000$ agents), $\e_{P} = 0.05$ and $\e_{D} = 0.05$. Agents' preferences $\vec{\xi}_{\mu}$ are sampled from a standard normal distribution and the initial distribution of the prices is taken from the uniform distribution $[0, 2]$, such that the average is one\footnote{Our simulations were in fact done with the lower bound on prices i.e. $x_{m}=0.01$ and not zero. This has no material impact on the present results however.}.
 
We further fix $\omega=0.2$ and $\gamma=1.0$.
We have explored numerically other parameter values. In fact, most of these parameters turn out to play a minor role and do not change the qualitative behaviour of the model, except for the level of allowed debt $\sigma$, which determines the emergent behaviour of our toy economy. This observation dovetails with the results reported in \cite{Gualdi2015a}, where the maximum amount of indebtedness of firms before bankruptcy plays a major role as well.

\subsection{Role of the debt limit: Macro-level}
\begin{figure}[t]
\centering
\includegraphics[scale=0.85]{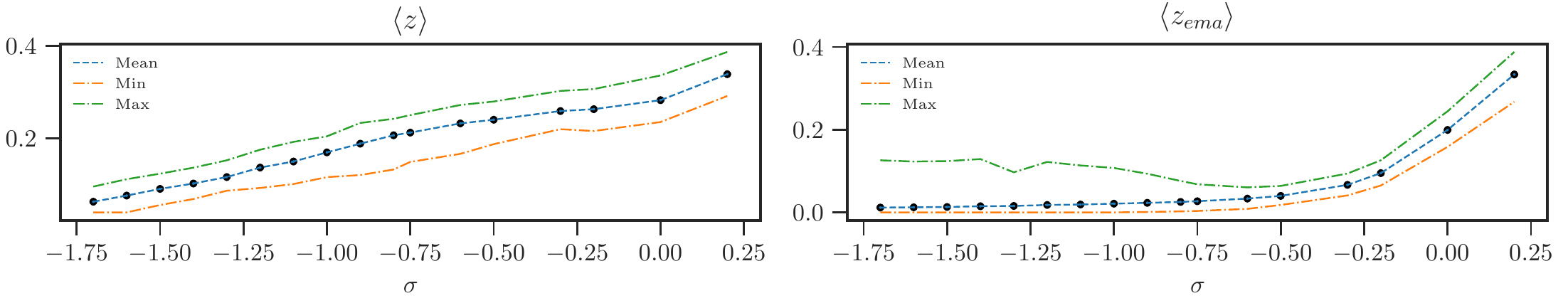}
\caption{Behavior of thermodynamic variables as a function of $\sigma$: $\langle z \rangle$ and $\langle z_{\textrm{ema}} \rangle$. Also represented are the maximum and minimum values attained during the simulation. }
\label{fig:thermo_1_fixed_omega}
\end{figure}

As a first step towards understanding the influence of $\sigma$ on the model, we choose a few macroscopic variables: 
\begin{enumerate}
\setlength{\itemsep}{1pt}
\item \(z_{\textrm{ema}}\) is defined as the fraction of agents who do not satisfy the time averaged constraint: 
$\pi_{\mu}^{\text{ema}}(t) < \sigma$;
\item $z$ is defined as the fraction of agents that do not satisfy the instantaneous constraint, Eq.~(\ref{define_h_mu}), at time $t$.
\end{enumerate}
As $\sigma$ is reduced (i.e. allowed debt becomes larger), the fraction of unsatisfied agents $z_{\textrm{ema}}$ and unsatisfied constraints in Eq.~(\ref{define_h_mu}) tend to zero, on average, see Figure~\ref{fig:thermo_1_fixed_omega}. This points to the interpretation that, as expected, an increased debt level lends flexibility to agents and permits them to live another day. However, beyond a certain negative value of $\sigma$ ($\approx -0.75$), we observe that the maximum value attained by $z_{\textrm{ema}}$ during the whole time series actually starts increasing, see Figure~\ref{fig:thermo_1_fixed_omega}-b. 
\medskip

\begin{figure}[H]
\includegraphics[width=\textwidth]{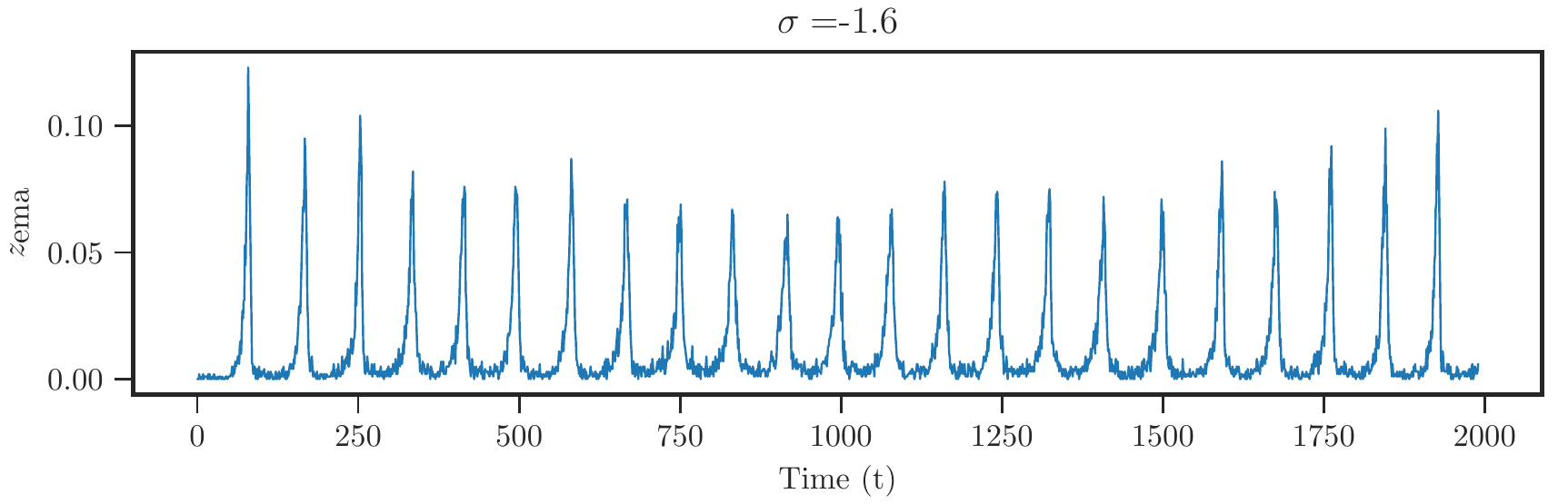}
\includegraphics[width=\textwidth]{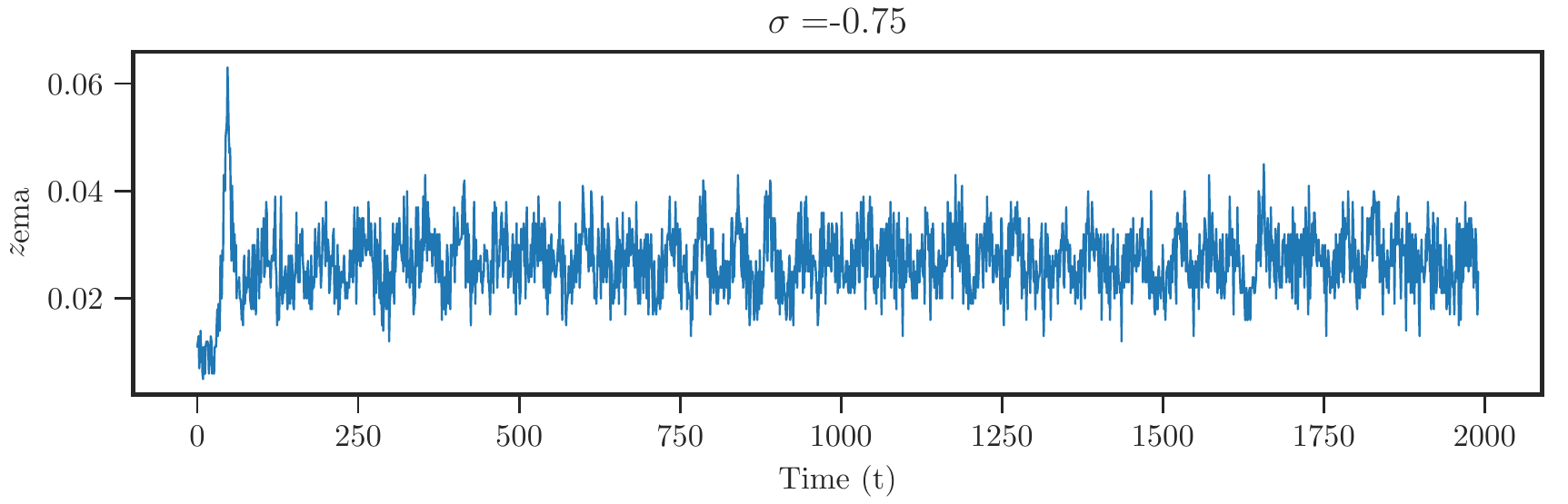}
\includegraphics[width=\textwidth]{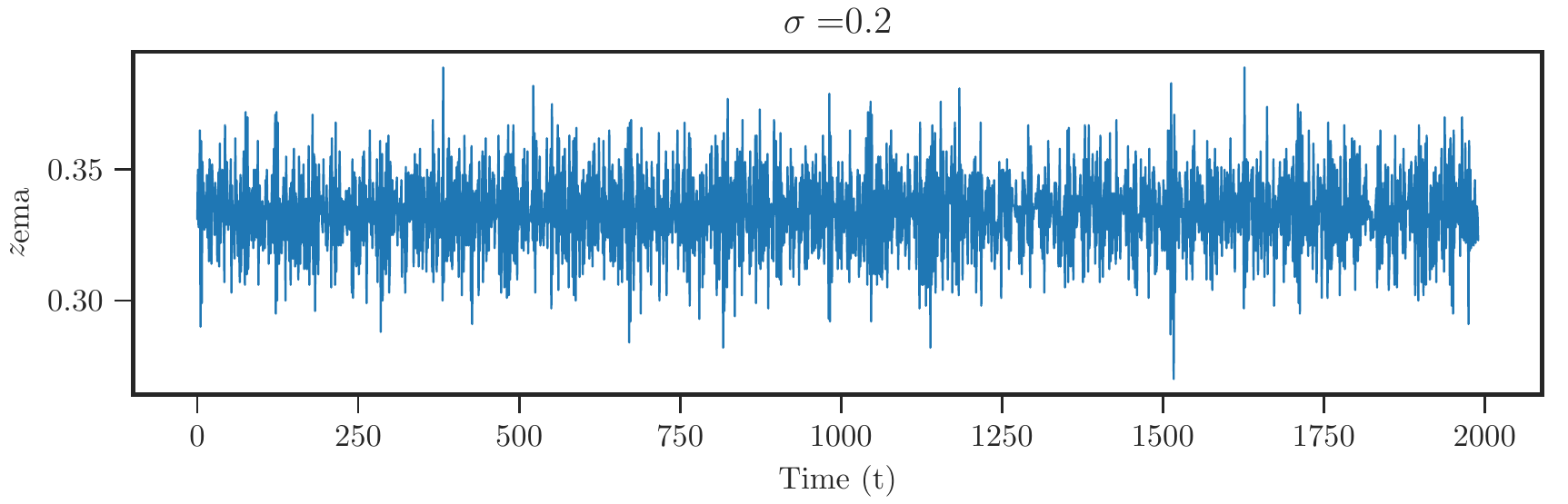}
\caption{Variation of the fraction of unsatisfied agents $z_{\textrm{ema}}$ as a function of time for three values of $\sigma$, in the EC (top), 
S (middle) and U (bottom) regimes. For $\sigma=-1.6$, the agents are removed and replaced periodically whereas for $\sigma=-0.75$, periodicity is lost. In the bottom panel, the $\sigma=0.2$ case corresponding to high-profit is shown.}
\label{fig:dynamics_zema}
\end{figure}

To better understand why the maximum of $z_{\textrm{ema}}$ increases beyond a certain value of $\sigma$, we plot in Figure~\ref{fig:dynamics_zema} the time series for $z_{\textrm{ema}}$ for three values of $\sigma$. 
Three distinct behaviors are observed: for positive, or mildly negative values of $\sigma$, the default rate is large. For $\sigma \sim 0.2$, at every time step close to 30\% of the agents are being removed at every step. As $\sigma$ is reduced, this default rate goes down and for $\sigma=-0.75$, a small fraction of the agents are removed at any time, about 3\%. On reducing the value of $\sigma$ further ($\sim -1.6$), we observe a clear periodicity in the time series of $z_{\textrm{ema}}$, with the appearance of regular spikes (which persist forever in the simulation). At the peaks, close to 10\% of the agents are removed from the economy. These oscillations in the rate of bankruptcies is reminiscent of a similar effect reported in \cite{Gualdi2015a} and explained in \cite{Gualdi2015b}. We will see below that a similar mechanism is at play here as well.  
\medskip

This suggests the existence of three regimes: 
\begin{enumerate}
    \item Endogenous Crises (EC): this occurs for small values of $\sigma$ (high levels of allowed debt), $\sigma \lesssim -0.75$, where periodic spikes of bankruptcies are observed; 
    \item Stable Phase (S): this occurs for an intermediate range of debt, $-0.75 \lesssim \sigma \lesssim -0.25$, where the economy reaches a stationary state characterized by a few bankruptcies;
    \item Unstable phase (U): this occurs for  positive values of minimal profits $\sigma$, where the economy features a high rate of bankruptcies.
\end{enumerate}

The behavior for intermediate (S) and low (EC) levels of debt is reminiscent of the self-planting phenomenon discovered in~\cite{Sharma2019} for a very similar model. In that case, below some critical value $\sigma_{c}$ the dynamics was able to find a configuration where all constraints (agents) were satisfied, while above $\sigma_{c}$ the dynamics reached a steady-state value for the fraction of unsatisfied constraints. 
Differently from~\cite{Sharma2019}, however, in the present context the dynamics is not halted since agents can continue to participate in transactions and update their preferences according to the movement of the prices.

\begin{figure}[H]
\centering 
\includegraphics[scale=0.7]{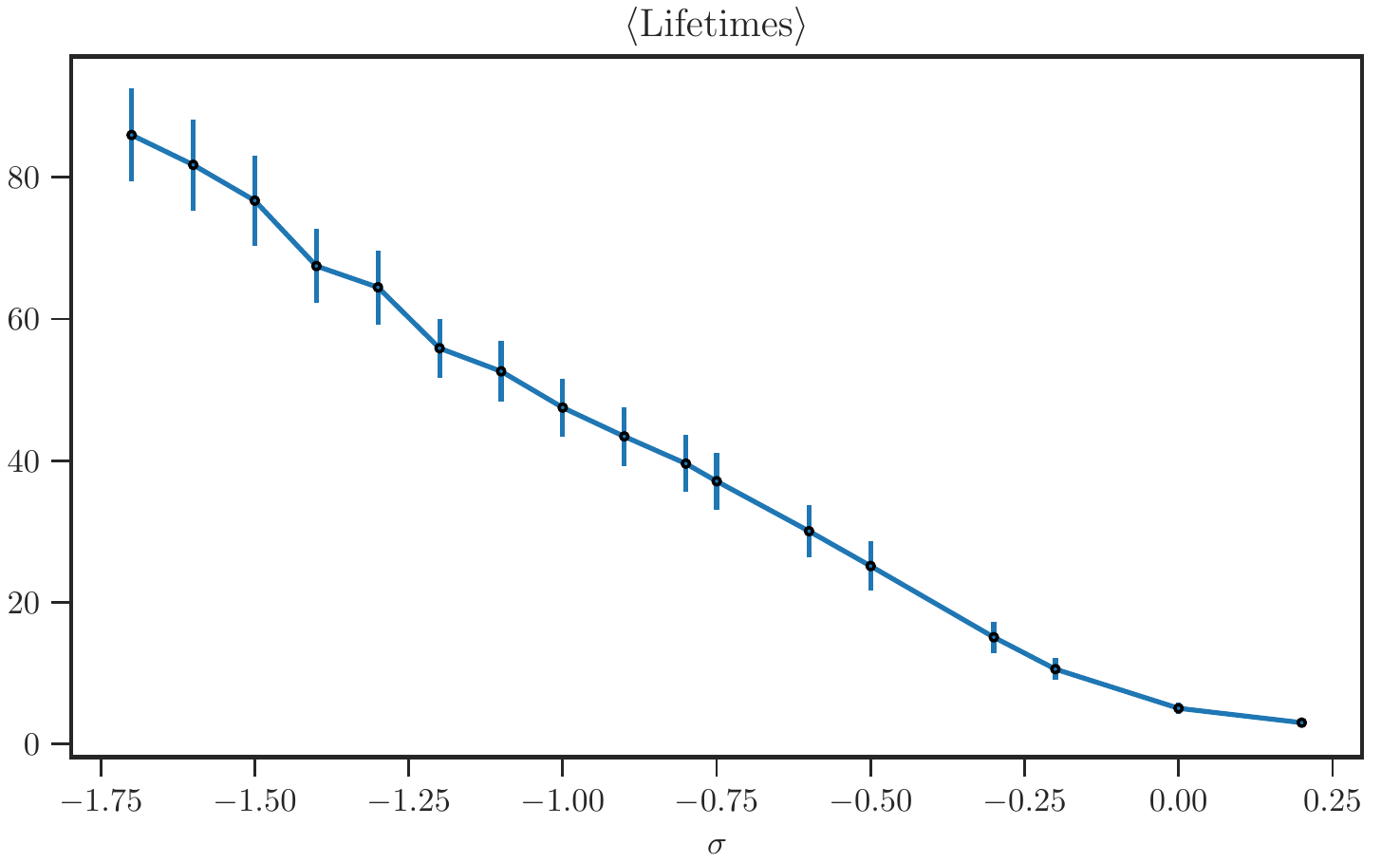}
\caption{Dependence of agent lifetimes as a function of $\sigma$. }
\label{fig:lifetimes_variation}
\end{figure}

We also measure 
the average number of time-steps during which an agent participates in the economy. 
The results are shown in Figure~\ref{fig:lifetimes_variation}, where we show how agent lifetimes change and the average number of times an agent is removed as a function of $\sigma$.
It is natural to expect that the lifetime of agents is higher when $\sigma$ is small, since it corresponds to a situation where $z_{\textrm{ema}}$ is low on average. Fewer agents are removed in each time-step in this regime (EC or S) and hence agents participate and live longer. The opposite situation is produced when there is no possibility of debt leading to the frequent removal of a large fraction of agents which in turn leads to smaller agent lifetimes. The U regime is characterized by the fact that agents live for not more than 3-4 time-steps, i.e. the timescale of the exponential moving average $\sim 1/\omega = 5$. On the contrary, the lifetimes of agents in the EC or S regimes are much larger than the averaging timescale. In the EC regime, the agent lifetime is comparable to the period of the ``business cycle'' (i.e. the period in $z_{\textrm{ema}}$ oscillations), which can be of the order of hundreds of time-steps. If a ``time-step'' is 3 months, this corresponds to an economic activity lifetime of 25 years.

\begin{figure}[H]
\centering
\includegraphics[width=\linewidth]{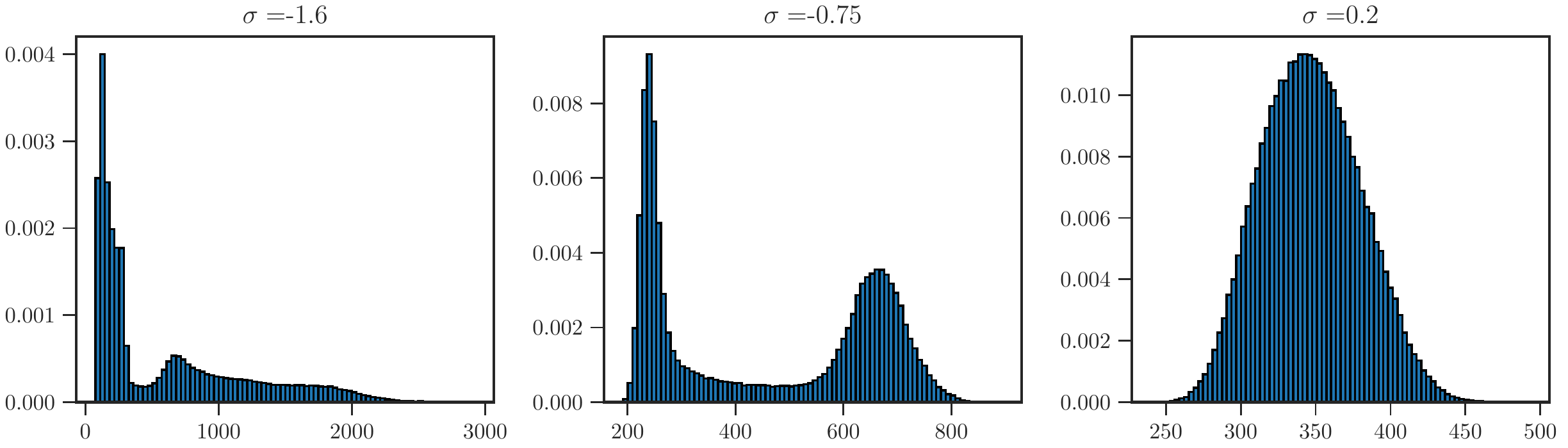}
\caption{The demand distributions for different values of $\sigma$.  We note that in the S regime (middle), the distribution is bimodal, while in the U regime (right) the distributions remain quasi-Gaussian. In the EC regime (left), the distribution has a long tail. These distributions are obtained by taking the distribution of the complete demand time series over all goods. }
\label{fig:dynamics_demand_distrib}
\end{figure}

Having considered how macro indicators behave as a function of $\sigma$, we now study how $\sigma$ influences levels of supply and demand. The EC and S regimes are characterized by agents persisting for long times with only a few agents being replaced at any given time step (except during crises in the EC regime). Longer living agents influence the distributions of the goods being exchanged in the economy. We show how the demand distributions vary as we tune $\sigma$ in Figure~\ref{fig:dynamics_demand_distrib}.
We observe that for intermediate values of $\sigma=-0.75$ in the S regime, the distributions is bimodal. The bimodality persists for a lower value of $\sigma$ in the EC regime, with the second peak presenting a heavy right tail. For positive $\sigma$ in the U regime, on the other hand, the distribution remains Gaussian, since agents' preferences are initially sampled from a normal distribution, and the 
process that updates agents' preferences (Eqs~\ref{def:update1},\ref{def:update2}) does not have time to operate before agents are replaced.

\begin{figure}[!h]
\centering
\includegraphics[width=.9\linewidth]{figures/scatter_prices_supplies_demands.pdf}
\caption{Scatter plot of supply and demand versus the prices of goods. In the first column, we have $\sigma=-1.6$ with supplies and demands concentrated at the extremes, corroborating the bimodal nature of their distributions. The intermediate case with $\sigma=-0.75$ is similar with some goods' supplies occurring between the two extremes. The last column shows the case for $\sigma=0.2$ where the supplies and demands are uncorrelated with the prices of the goods. The color map corresponds to the number of points in a given region.}
\label{fig:dependence_supply_prices}
\end{figure}

\begin{figure}[t]
\centering 
\includegraphics[width=\linewidth]{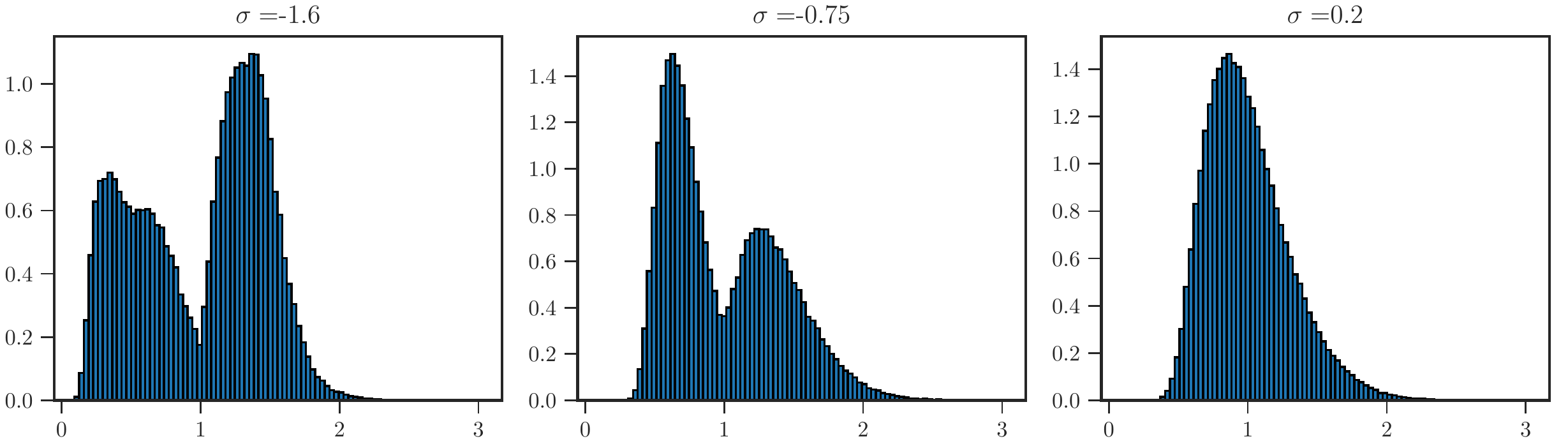}
\caption{Price distributions for different values of $\sigma$, as in Figure~\ref{fig:dynamics_demand_distrib}. These distributions are computed over the complete time-series of the prices of goods. }
\label{fig:price_distrib}
\end{figure}

The rules governing agents' behaviour also couple the level of supply and demand to the price of goods.  In general, we expect that cheaper goods will be in higher demand (and hence supply) and vice-versa for more expensive goods. We test this hypothesis in
Figure~\ref{fig:dependence_supply_prices},
where we show scatter plots of supplies and demands versus prices.
We find that, as expected, both demand and supply of goods are anti-correlated with their price in the EC and S regimes.  In the U regime, instead, such behavior is not observed: the process of price formations is completely random because agents fail before being able to adapt their preferences.
Figure~\ref{fig:dependence_supply_prices} also shows the signature of the bimodality of the demand and supply distributions in the EC and S regimes. This is confirmed by the observation that the price distribution is bimodal as well,
as shown in Figure~\ref{fig:price_distrib}. 

Hence, our model generates three well distinct regimes.
The U regime is somehow pathological: agents are replaced immediately after their introduction, hence their preferences are completely random and unable to adapt, and prices are also random, as a consequence.
On the contrary,
in the EC and S regimes, agents remain in the economy long enough to be able to adapt their prices. In the EC regime, we observe periodic spikes of bankruptcies during which all agents are replaced; the agents' lifetime is then comparable to the periodicity of crises. In the S regimes, agents are replaced at a constant but very low rate. In both cases,
one observes what could be called a ``speciation'' of goods. While all goods are {\it a priori} equivalent, the system self-organizes in such a way to create two categories of goods: cheap goods in high demand on the one hand, and expensive goods in low demand on the other. Note however that since this speciation is endogenous to the dynamics, it is also temporary: goods switch from one group to the other with time. The details of this dynamical process is what we examine next.

\subsection{Role of the debt limit: Dynamics}

The study of macroscopic observables 
has demonstrated that $\sigma$ is a key control parameter, which drives the system through 
three distinct regimes. We now turn to an analysis of the influence of $\sigma$ on the dynamics of the economy. 

\begin{figure}[t]
\includegraphics[width=\textwidth]{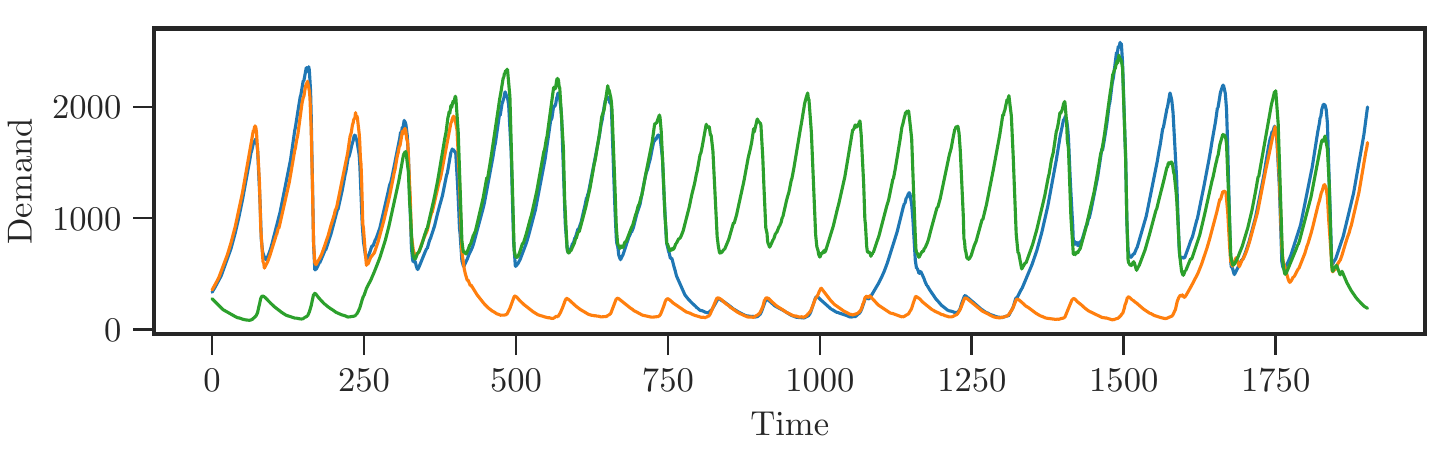}
\includegraphics[width=\textwidth]{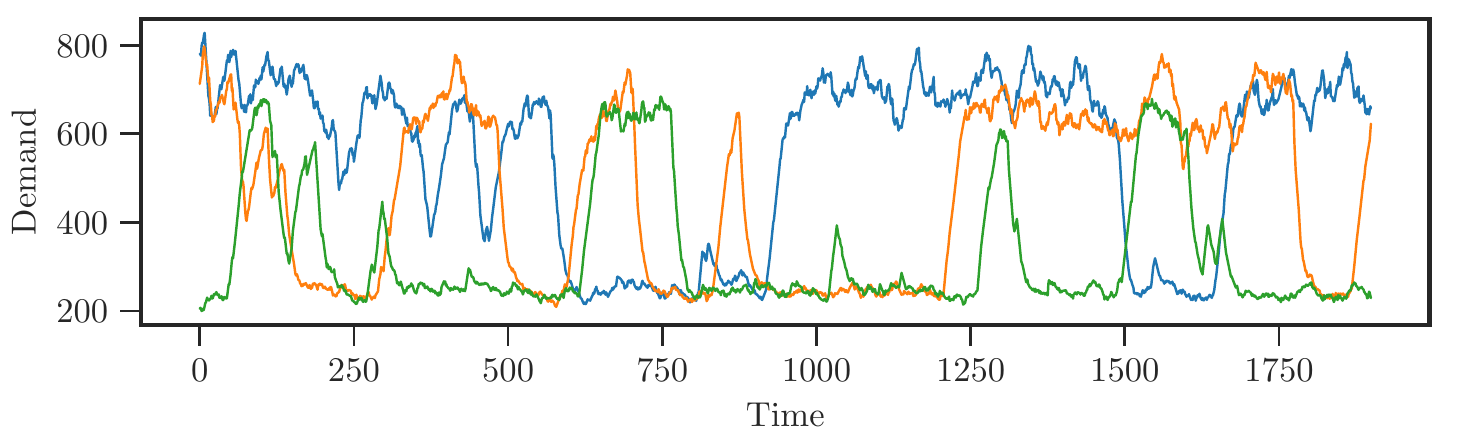}
\includegraphics[width=\textwidth]{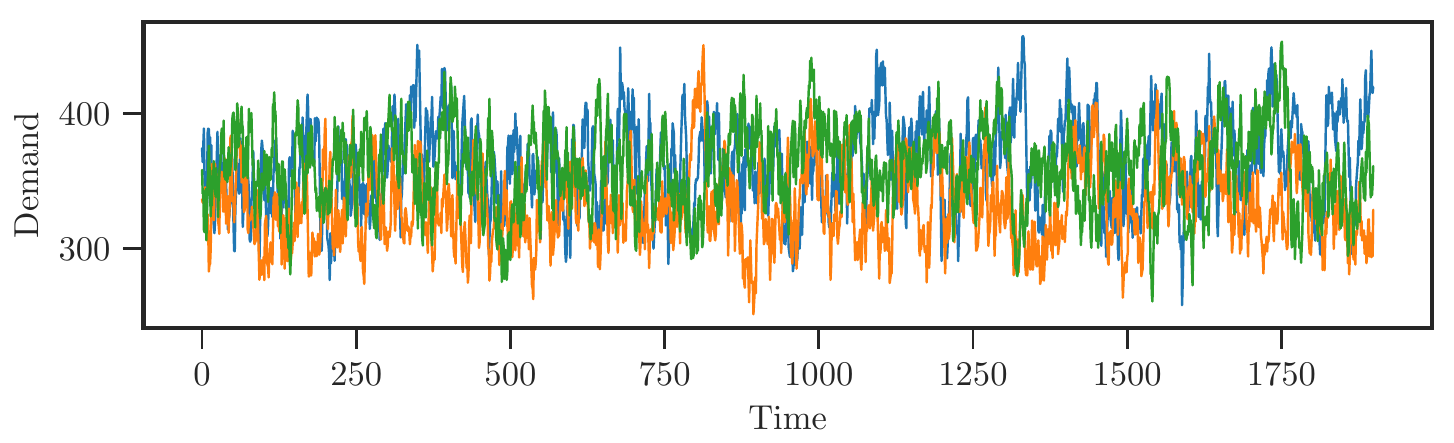}
\caption{Time series of the demand of three random goods for three values of $\sigma$. Top: $\sigma=-1.6$ (EC regime) where we have cyclical rise and fall in demand. Middle: $\sigma=-0.75$ (S regime) where goods can switch from having high demand to having low demand (and vice-versa). As one good demand falls, another low-demand good takes over. Bottom: $\sigma=0.2$ (U regime) where no coherent trend is found.  }
\label{fig:dynamics_switching}
\end{figure}

We begin by observing how the demand of individual goods varies with time in these three regimes. 
In Figure~\ref{fig:dynamics_switching} we show time-series of the demand of three randomly chosen goods for three values of $\sigma$. 
In the EC regime, we observe that the demand level of the goods is also periodic. This is a consequence of the existence of periodic crashes: as a large number of agents are periodically removed, the corresponding demand for goods also undergoes periodic swings. Interestingly, goods with low demand are out of phase with goods with high demand. 
In the U regime ($\sigma=0.2$), the demand for goods fluctuates around an average value, within $\sim \pm 10\%$. This follows the behaviour observed in the $z_{\textrm{ema}}$ time-series as well: a significant proportion of agents is removed at each time step and hence the level of demand remains stable around its average. 
For the intermediate S regime ($\sigma = -0.75$), the situation is quite complex. The demand for individual goods shows large variations: one good might start with high demand before being replaced by another good, within rather short time intervals. This suggests that the goods that are present in the two peaks of the bimodal distribution of demands (Figure~\ref{fig:dynamics_demand_distrib}) are not the same but keep changing in time. 
We indeed observe that goods keep switching from the right peak (high demand) to the left peak (low demand) dynamically, while maintaining the bimodal character of the distributions globally intact. A similar scenario with endogenous switches is found in ~\cite{Yanagita2010}: firms compete among themselves to gain market share with boundedly rational consumers choosing firms to maximise their utility. This interaction leads to firms dynamically exchanging positions of monopoly with each other.  

\begin{figure}[t]
    \includegraphics[width=\linewidth]{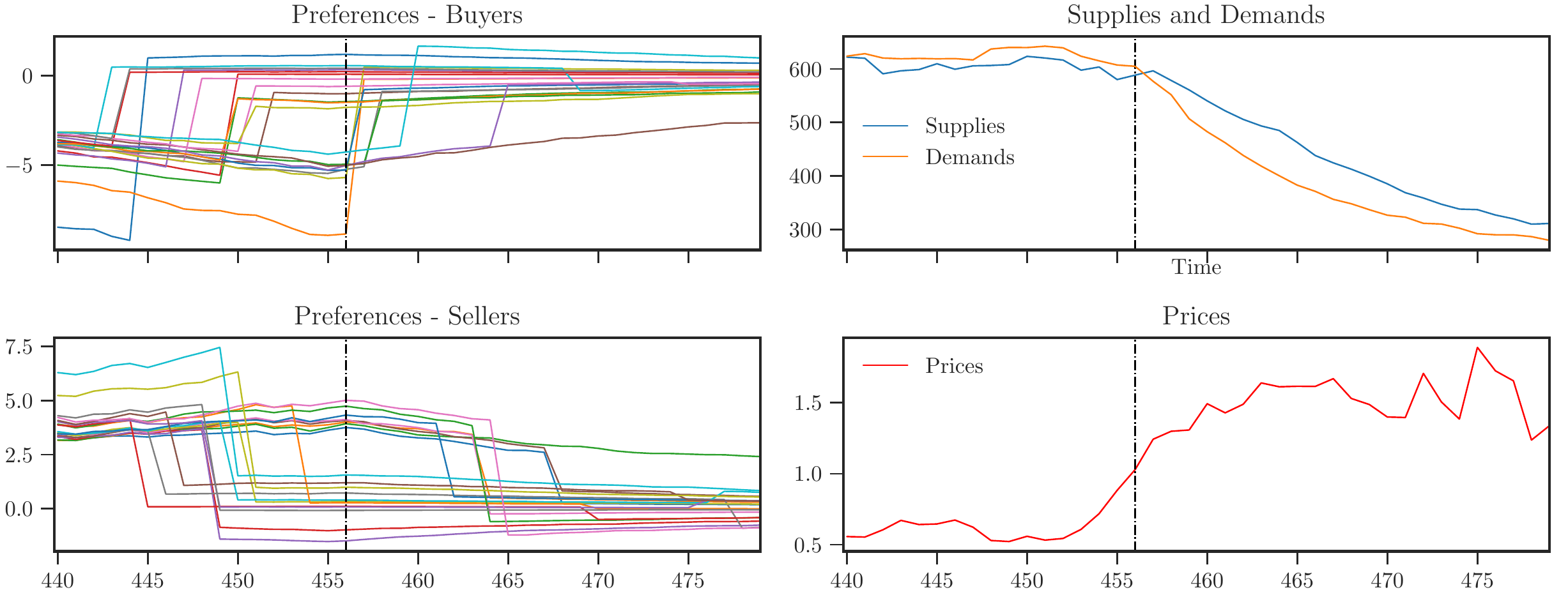}
    \caption{Understanding the switch from high to low demand of a good. {\it Left column}: The individual trajectories of the preferences of agents who are buyers (top) and sellers (bottom) is shown. The removal of an agent is shown by an abrupt jump or fall in the preference. {\it Right column}: The total supply and demand for the good (top) and its price (bottom) are shown. The black dotted line is the point when $p_{i} >1$.}
    \label{fig:preference_comparison_dynamics}
\end{figure}

As a means of understanding the dynamical switching between goods in the EC regime, we consider the case of a particular good whose demand (and hence supply) falls. A preliminary observation we make is that the behavioral rules for the agents imply that as soon as the price of a good becomes high (greater than the average price of goods, equal to 1 here), then agents will reduce their demand. Hence, we expect that as soon as the price of a good reaches 1, we will observe a fall in demand, which will lead to a fall in supply. The fall in supply in turn will produce a further rise in prices since sellers will seek to maintain their previous profit levels at reduced demand. This increase in price feeds back into suppressed demand and the feedback loop leads to a rapid collapse. This situation is indeed borne out in the data as shown in Figure~\ref{fig:preference_comparison_dynamics}. In the left column, we show how agents reduce their demand and how supply follows in lockstep. At the aggregate level, we observe the rapid fall of the supply and demand of the good as soon as $p_{i} >1$ (right column top). 

Still, two questions remain open.
The first question is: what factors lead to the initial price increase of the good itself? We observe in Figure~\ref{fig:preference_comparison_dynamics} that the price continues to increase for a certain number of time-steps before reaching $p_{i}=1$. This increase in price can be understood by the failure of big buyers of the good. Producers of this good then face reduced demand and hence must lower production according to Eq.~(\ref{def:update1}). The prices then have to adjust upwards since producers have to satisfy their budget constraint at a lower scale of production. The second question is: what precipitates the failure of a big buyer or big seller? To understand this better, we define a quantity called the $f$-index:
\begin{align}
\begin{split}
    f_{i}^{\textrm{sellers}}(t) &= \sum_{\mu} \xi_{\mu}^{i}(t) \, \Theta(\xi_{\mu}^{i}(t))\, \, \Theta\big(\sigma-\pi_{\mu}^{\textrm{ema}}(t+1)\big) \, ,\\
     f_{i}^{\textrm{buyers}}(t) &= \sum_{\mu} | \xi_{\mu}^{i}(t) | \, \Theta(-\xi_{\mu}^{i}(t)) \, \, \Theta\big(\sigma-\pi_{\mu}^{\textrm{ema}}(t+1)\big) \, .
\end{split}
\label{eq:define_f_index}
\end{align}

\begin{figure}[!h]
\centering
\includegraphics[width=\linewidth]{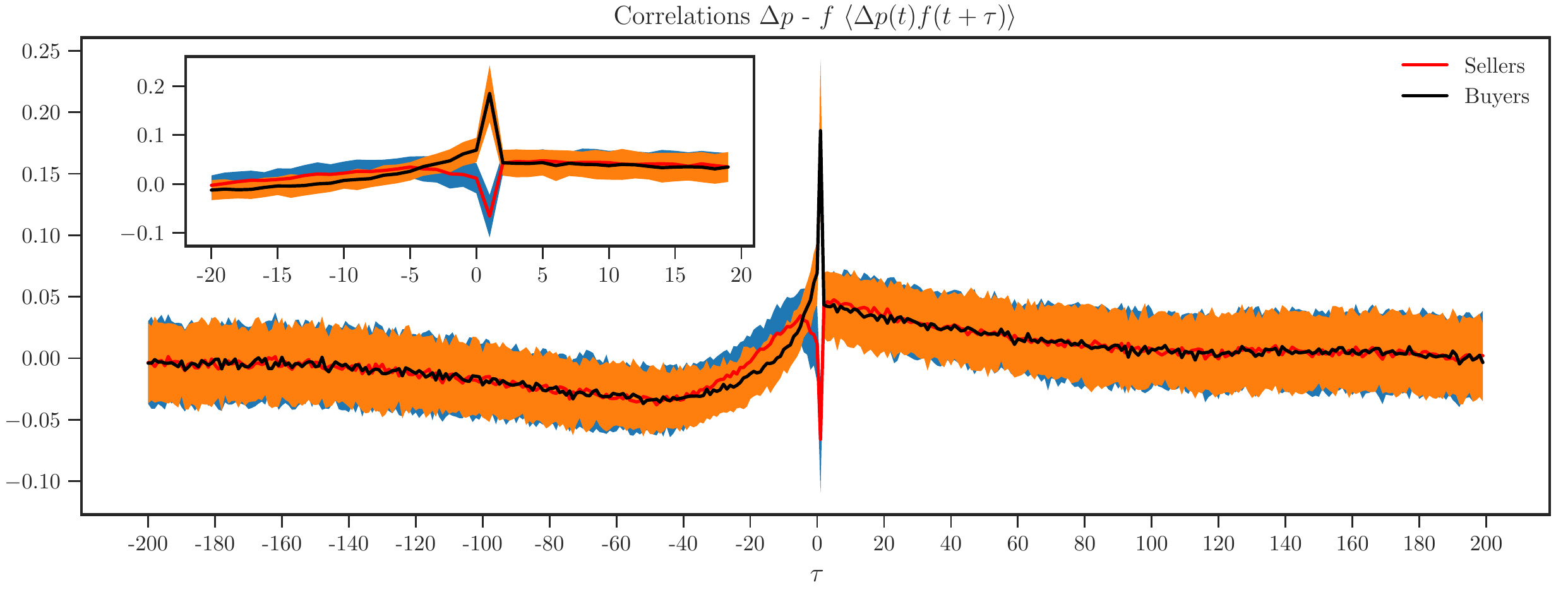}
\caption{ Correlation between changes in prices $\Delta p$ with the $f$-index. The correlation shows a sharp peak at $\tau=1$. Inset shows a zoom near the peak of the correlation function. The correlation here is $\langle \Delta p(t) f(t+\tau) \rangle$ averaged over all the goods. }
\label{fig:z_index_correlations}
\end{figure}
Thus the $f$-index computes the decrease in supply or demand for a good due to the agents who will go bankrupt in the next step. We measure the correlation of the $f$-index with the change in prices. The average of this correlation over all goods is shown in Figure~\ref{fig:z_index_correlations}. We observe that for both buyers and sellers the correlations are peaked at $\tau=1$. The positive correlation observed for buyers suggests that an increase in the price of goods is accompanied with the removal of buyers in the next time step. On the other hand, the correlation for sellers is negative implying that a reduction in the prices corresponds to the sellers being removed. 

We thus conclude that purely random fluctuations in the price of a good can engender, through a feedback loop, the failure of both sellers and buyers and produce the peculiar dynamics observed in the middle panel of Figure~\ref{fig:dynamics_switching}. One might argue that this is due to the myopic nature of our agents and the strategy they use. This might well be; on the other hand, it is dangerous to bet the stability of the economy on the purported rationality of agents.  

\begin{figure}[t]
\centering
\includegraphics[width=.96\textwidth]{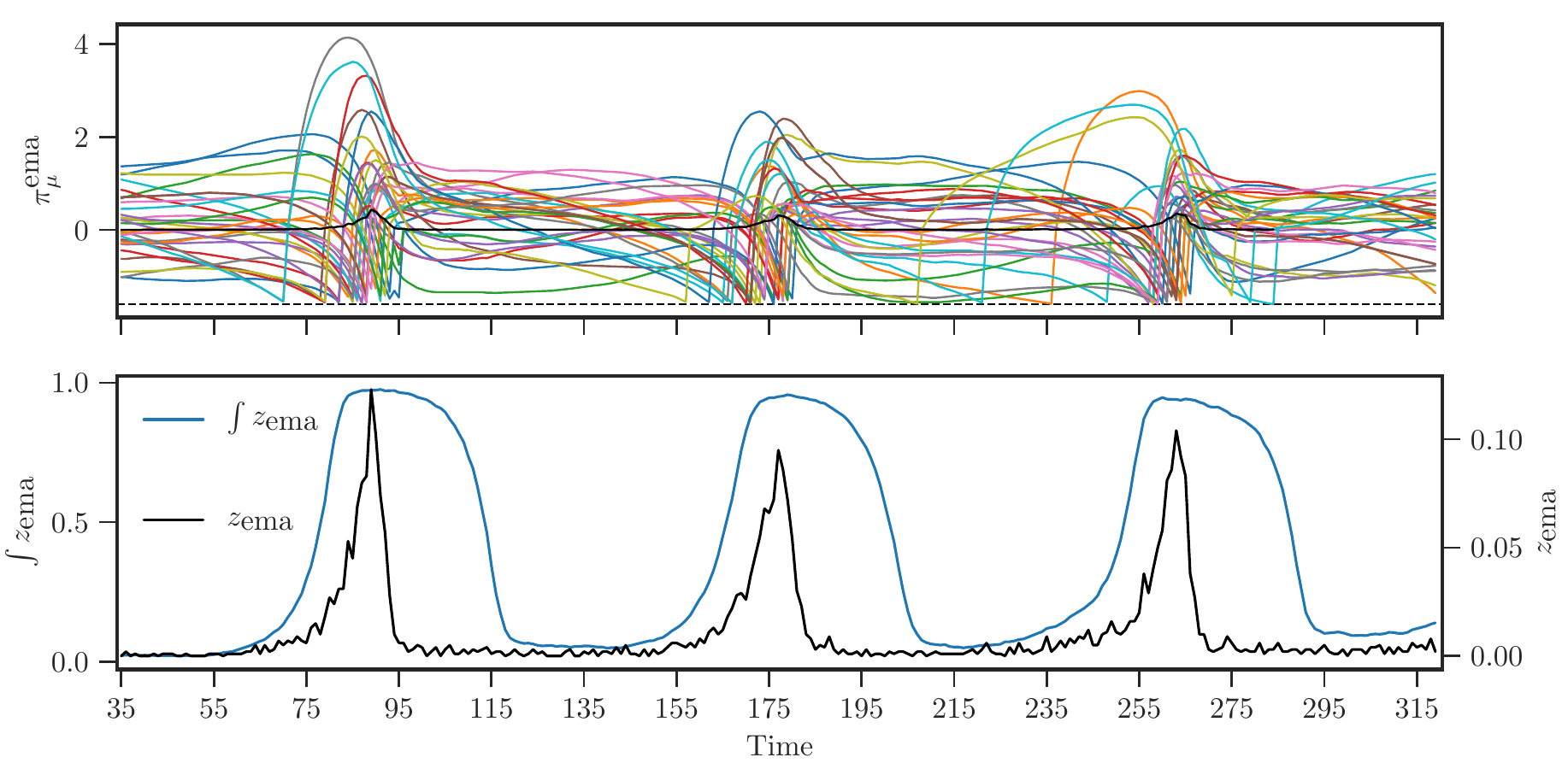}
\caption{{\it Top}: $\pi_{\mu}^{\textrm{ema}}$ time-series for $\sigma=-1.6$ is shown for 20 agents. The dark black line is the average value of $\pi_{\mu}$ at which agents are re-injected. {\it Bottom}: $z_{\textrm{ema}}$ (in black) with a moving integral of $z_{\textrm{ema}}$ showing that during the crash almost all agents are removed and replaced. }
\label{fig:dynamics_piema_cycle}
\end{figure}

We now move to the dynamics of individual agents. For sufficiently small values of $\sigma$, in the EC regime, we have observed that $z_{\textrm{ema}}$ is cyclical: a significant fraction of the agents are removed during crashes, which cover about 10-20 time-steps separated by periods with low removal rates. In the top panel of Figure~\ref{fig:dynamics_piema_cycle} we show the $\pi_{\mu}^{\textrm{ema}}$ time-series for 20  randomly chosen agents. We observe that those agents who are well above their threshold and with disparate budgets are nevertheless removed together during a crash. Furthermore, the lower panel of Figure~\ref{fig:dynamics_piema_cycle} shows that over one period of the cycle almost all the agents are removed and replaced.  

\begin{figure}[t]
\centering
\includegraphics[width=\linewidth]{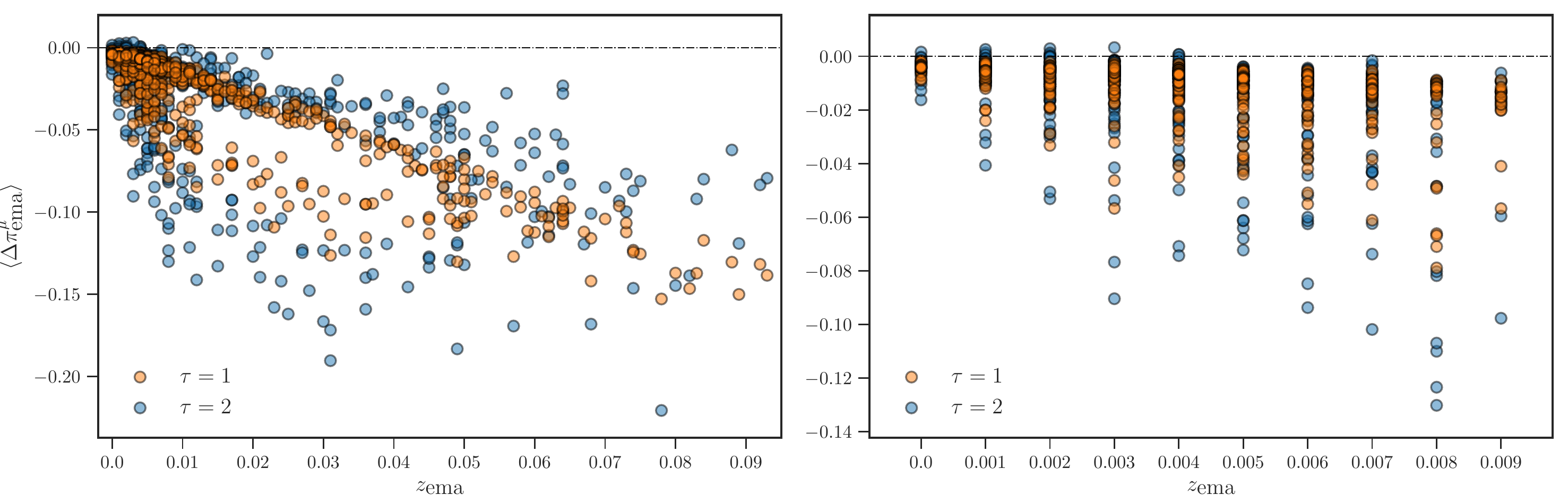}
\caption{Variation of the change in the profits $\pi_{\mu}^{\textrm{ema}}$ averaged over all remaining agents following the removal of an agent plotted against $z_{\textrm{ema}}$. Two distinct cases are shown, $\tau=1$ and $\tau=2$ corresponding to changes in the profit of surviving agents one time-step or two time-steps after the removal of agents. {\it Left}: $\Delta \pi^{\mu}_{\textrm{ema}}$ against $z_{\textrm{ema}}$ computed over the complete evolution of the economy. {\it Right}: Zoom of $\Delta \pi^{\mu}_{\textrm{ema}}$ for small values of $z_{\textrm{ema}}$.}
\label{fig:delta_pi_ema_after_removal}
\end{figure}

The mechanism leading to the cyclical behavior can be understood via the feedback that exists between the failure of agents and the profit trajectories of the agents. As discussed in~\cite{Gualdi2015b}, a biased random walk along with an absorbing boundary condition and re-injections can under certain conditions lead to synchronized crisis waves. In our case, we observe that the removal of an agent produces a small but systematic reduction in $\pi_{\mu}^{\textrm{ema}}$ for all the other surviving agents -- i.e. the failure of one agent fragilises the rest of the community. This is exactly the mechanism at the origin of synchronisation discussed in ~\cite{Gualdi2015b}, leading to periodic crises of fully endogenous nature.

In order to provide support to this interpretation, in Figure~\ref{fig:delta_pi_ema_after_removal}, we plot the change in profits $\Delta \pi_{\textrm{ema}}^{\mu}=\pi^{\mu}_{\textrm{ema}}(t+\tau) - \pi^{\mu}_{\textrm{ema}}(t)$ against the fraction of agents removed $z_{\textrm{ema}}$ at each time-step. This shows that the average (over surviving agents) change in profits is nearly always negative for the surviving agents and this is true irrespective of the number of agents removed, even when small (see Figure~\ref{fig:delta_pi_ema_after_removal}, right). It is quite satisfying to see that the mechanism at play in a very different ABM, namely ``Mark0'' \cite{Gualdi2015a}, appears in another guise in the present context. 


\section{Conclusion} 

In this work we presented a prototype agent-based model with a budgetary constraint at its core. Our basic assumption is that preferences (i.e. supply and demand) cannot adapt immediately, but only do so with a lag. The market, on the other hand, provides the best set of prices given the budget constraints. These budgetary constraints lead us to interpret the economy as a constraint satisfaction problem (CSP) -- given agents' preferences and their budgets, what configuration of prices achieves maximal ``satisfaction''? 

We restricted in this paper to the simplest case where the global constraint on prices is linear, fixing the average price to unity. This means that the CSP is, at each time step, convex and its solution unique. Other, non-linear constraints could be envisaged and would probably generate an even richer phenomenology due to the existence of an exponentially large number of equilibria~\cite{Franz2017a,Sharma2019}, see also~\cite{galluccio1998rational} in a related context. But even the simplest specification of the model leads to very interesting (and sometimes unexpected) features, such as spontaneous speciation of goods, or waves of synchronized bankruptcies.  

We have found that the model exhibits three regimes as a function of the amount of debt that each agent can accumulate before defaulting. At high debt (strongly negative $\sigma$), we observe the appearance of endogenous cycles during which the economy goes from a state of stability with very few bankruptcies to a state with a high bankruptcy rate. Lowering the allowed debt (increasing $\sigma$) takes the model to a regime where the economy is stable with very low bankruptcy rate and no aggregate-level crises. In this regime, the system self-organizes in such a way to create two categories of goods: cheap goods in high demand on the one hand, and expensive goods in low demand on the other. Goods switch from one category to the other with time. These switches are triggered by random fluctuations, but when they occur, a feedback loop accelerates the transition and generates abrupt ``crashes'', when favored products quickly lose their luster. 

Finally, there is a low debt or high required profit regime (positive $\sigma$) where the bankruptcy rate is extremely high with a large fraction of the agents going bankrupt at every time step. In other words, in this regime agents never have time to adapt and the economy remains structure-less. 

We have shown that the emergence of waves of defaults in the endogenous crisis regime is triggered by the same mechanism as the one proposed in \cite{Gualdi2015b}, namely the degradation of the balance sheet of surviving agents when one agent goes under. This mechanism is, we believe, very general and is likely to be present in many real world situations (see \cite{Gualdi2015a, Gualdi2015b} and \cite{Ledger2018} for recent mathematical developments. See also~\cite{Pangallo2020} where a more general framework, inspired by the literature on dynamical systems, has been proposed to understand synchronization within endogenously created business cycles).


Following the crisis of 2008, the issue of credit expansion  and its effects on the economy has found center-stage in macroeconomics. There is now increasing evidence that financial crises are preceded by protracted periods of credit expansion.  Empirically, it has been observed that the higher the private debt buildup, the deeper is the downturn~\cite{Jorda2011}. Both ABM and standard DSGE approaches have sought to understand the details of the leverage cycle: Ref.~\cite{Adrian2012} studies the leverage of financial entities intermediating between households and industry. They conclude that increased leverage does lead to increased output but at the cost of systemic risk. Aymanns and co-authors build an agent-based model to uncover the dynamics of the leverage cycle based on the risk perception of actors like banks~\cite{Aymanns2015, Aymanns2016}. They propose a detailed study underlying the importance of managing leverage systematically to dampen endogenously created debt-driven boom-bust cycles. Importantly, they find sustainable levels of leverage exceeding which the dynamics becomes unstable and chaotic. 

Our present model hence confirms the central role that debt levels play in the stability of the economy:  too high a debt level  and  we  have  periodic  crises,  too  low  a  debt  level  and  the  agents  cannot  sustain  themselves  long  enough  for long-lived structures to appear and survive. Our work presents a break from previous studies on the leverage cycle by coupling the production output and trading within the economy with agents' budgetary constraints. Since agents in our model get credit for free, an interesting direction for future exploration is the introduction of a bank (or banking sector) to set a price for borrowing through the interest rate.

Let us reiterate our belief that the CSP paradigm will, in due course, play an important role in economics. The idea is that interacting constraints (which could be seen as the defining feature of any economic system) generate complex, non-trivial phenomena at the aggregate level. Within our framework, budgetary constraints --- a staple of most ABMs --- are local, agent-specific constraints, which aggregate upscale and set economy-wide prices. Budgetary constraints also couple the global production level to agents' debt in a straight-forward fashion. Moreover, while other approaches towards leverage dynamics rely upon further behavioral assumptions (for instance in~\cite{Seppecher2015} an agent-based model is studied with supplementary opinion dynamics leading to changes in macroeconomic regimes), within our model business cycles are produced through a purely market-driven feedback loop. 

Finally, CSPs like the perceptron are well-studied analytically. Hence integrating CSPs within an agent-based model also opens up the possibility of analytically understanding such models. 





\acknowledgments

DS would like to thank Jose Moran for illuminating discussions and feedback. 

\bibliographystyle{apsrev4-1}
\bibliography{references.bib}

\begin{thebibliography}{50}%
\makeatletter
\providecommand \@ifxundefined [1]{%
 \@ifx{#1\undefined}
}%
\providecommand \@ifnum [1]{%
 \ifnum #1\expandafter \@firstoftwo
 \else \expandafter \@secondoftwo
 \fi
}%
\providecommand \@ifx [1]{%
 \ifx #1\expandafter \@firstoftwo
 \else \expandafter \@secondoftwo
 \fi
}%
\providecommand \natexlab [1]{#1}%
\providecommand \enquote  [1]{``#1''}%
\providecommand \bibnamefont  [1]{#1}%
\providecommand \bibfnamefont [1]{#1}%
\providecommand \citenamefont [1]{#1}%
\providecommand \href@noop [0]{\@secondoftwo}%
\providecommand \href [0]{\begingroup \@sanitize@url \@href}%
\providecommand \@href[1]{\@@startlink{#1}\@@href}%
\providecommand \@@href[1]{\endgroup#1\@@endlink}%
\providecommand \@sanitize@url [0]{\catcode `\\12\catcode `\$12\catcode
  `\&12\catcode `\#12\catcode `\^12\catcode `\_12\catcode `\%12\relax}%
\providecommand \@@startlink[1]{}%
\providecommand \@@endlink[0]{}%
\providecommand \url  [0]{\begingroup\@sanitize@url \@url }%
\providecommand \@url [1]{\endgroup\@href {#1}{\urlprefix }}%
\providecommand \urlprefix  [0]{URL }%
\providecommand \Eprint [0]{\href }%
\providecommand \doibase [0]{http://dx.doi.org/}%
\providecommand \selectlanguage [0]{\@gobble}%
\providecommand \bibinfo  [0]{\@secondoftwo}%
\providecommand \bibfield  [0]{\@secondoftwo}%
\providecommand \translation [1]{[#1]}%
\providecommand \BibitemOpen [0]{}%
\providecommand \bibitemStop [0]{}%
\providecommand \bibitemNoStop [0]{.\EOS\space}%
\providecommand \EOS [0]{\spacefactor3000\relax}%
\providecommand \BibitemShut  [1]{\csname bibitem#1\endcsname}%
\let\auto@bib@innerbib\@empty
\bibitem [{\citenamefont {Vines}\ and\ \citenamefont
  {Wills}(2017)}]{Vines2017}%
  \BibitemOpen
  \bibfield  {author} {\bibinfo {author} {\bibfnamefont {D.}~\bibnamefont
  {Vines}}\ and\ \bibinfo {author} {\bibfnamefont {S.}~\bibnamefont {Wills}},\
  }\href {\doibase 10.1093/oxrep/grx062} {\bibfield  {journal} {\bibinfo
  {journal} {Oxford Review Of Economic Policy}\ }\textbf {\bibinfo {volume}
  {34}},\ \bibinfo {pages} {1} (\bibinfo {year} {2017})}\BibitemShut {NoStop}%
\bibitem [{\citenamefont {Blanchard}(2018)}]{Blanchard2018}%
  \BibitemOpen
  \bibfield  {author} {\bibinfo {author} {\bibfnamefont {O.}~\bibnamefont
  {Blanchard}},\ }\href {\doibase 10.1093/oxrep/grx045} {\bibfield  {journal}
  {\bibinfo  {journal} {Oxford Review of Economic Policy}\ }\textbf {\bibinfo
  {volume} {34}},\ \bibinfo {pages} {43} (\bibinfo {year} {2018})}\BibitemShut
  {NoStop}%
\bibitem [{\citenamefont {Sims}(1980)}]{Sims1980}%
  \BibitemOpen
  \bibfield  {author} {\bibinfo {author} {\bibfnamefont {C.}~\bibnamefont
  {Sims}},\ }\href {http://www.jstor.org/stable/1912017 .} {\bibfield
  {journal} {\bibinfo  {journal} {Econometrica}\ }\textbf {\bibinfo {volume}
  {48}},\ \bibinfo {pages} {1} (\bibinfo {year} {1980})}\BibitemShut {NoStop}%
\bibitem [{\citenamefont {Sargent}(1993)}]{Sargent1993}%
  \BibitemOpen
  \bibfield  {author} {\bibinfo {author} {\bibfnamefont {T.~J.}\ \bibnamefont
  {Sargent}},\ }\href {https://ideas.repec.org/b/oxp/obooks/9780198288695.html}
  {\emph {\bibinfo {title} {{Bounded Rationality in Macroeconomics: The Arne
  Ryde Memorial Lectures}}}}\ (\bibinfo  {publisher} {Oxford University
  Press},\ \bibinfo {year} {1993})\BibitemShut {NoStop}%
\bibitem [{\citenamefont {Shiller}(1981)}]{Shiller1981}%
  \BibitemOpen
  \bibfield  {author} {\bibinfo {author} {\bibfnamefont {R.~J.}\ \bibnamefont
  {Shiller}},\ }\href
  {http://ideas.repec.org/a/aea/aecrev/v71y1981i3p421-36.html} {\bibfield
  {journal} {\bibinfo  {journal} {American Economic Review}\ }\textbf {\bibinfo
  {volume} {71}},\ \bibinfo {pages} {421} (\bibinfo {year} {1981})}\BibitemShut
  {NoStop}%
\bibitem [{\citenamefont {Long}\ and\ \citenamefont
  {Plosser}(1983)}]{Long1983}%
  \BibitemOpen
  \bibfield  {author} {\bibinfo {author} {\bibfnamefont {J.~B.}\ \bibnamefont
  {Long}}\ and\ \bibinfo {author} {\bibfnamefont {C.~I.}\ \bibnamefont
  {Plosser}},\ }\href {\doibase https://doi.org/10.1086/261128} {\bibfield
  {journal} {\bibinfo  {journal} {Journal of Political Economy}\ }\textbf
  {\bibinfo {volume} {91}},\ \bibinfo {pages} {39} (\bibinfo {year}
  {1983})}\BibitemShut {NoStop}%
\bibitem [{\citenamefont {Bernanke}\ \emph {et~al.}(1996)\citenamefont
  {Bernanke}, \citenamefont {Gertler},\ and\ \citenamefont
  {Gilchrist}}]{Bernanke1996}%
  \BibitemOpen
  \bibfield  {author} {\bibinfo {author} {\bibfnamefont {B.}~\bibnamefont
  {Bernanke}}, \bibinfo {author} {\bibfnamefont {M.}~\bibnamefont {Gertler}}, \
  and\ \bibinfo {author} {\bibfnamefont {S.}~\bibnamefont {Gilchrist}},\ }\href
  {\doibase 10.2307/2109844} {\bibfield  {journal} {\bibinfo  {journal} {Review
  of Economics and Statistics}\ }\textbf {\bibinfo {volume} {78}},\ \bibinfo
  {pages} {1} (\bibinfo {year} {1996})}\BibitemShut {NoStop}%
\bibitem [{\citenamefont {Cochrane}(1994)}]{Cochrane1994}%
  \BibitemOpen
  \bibfield  {author} {\bibinfo {author} {\bibfnamefont {J.~H.}\ \bibnamefont
  {Cochrane}},\ }\href {\doibase 10.1016/0167-2231(94)00024-7} {\bibfield
  {journal} {\bibinfo  {journal} {Carnegie-Rochester Confer. Series on Public
  Policy}\ }\textbf {\bibinfo {volume} {41}},\ \bibinfo {pages} {295} (\bibinfo
  {year} {1994})}\BibitemShut {NoStop}%
\bibitem [{\citenamefont {Ma}(2018)}]{Ma2018}%
  \BibitemOpen
  \bibfield  {author} {\bibinfo {author} {\bibfnamefont {S.-K.}\ \bibnamefont
  {Ma}},\ }\href {\doibase https://doi.org/10.4324/9780429498886} {\emph
  {\bibinfo {title} {Modern Theory Of Critical Phenomena}}}\ (\bibinfo
  {publisher} {Routledge},\ \bibinfo {address} {New York},\ \bibinfo {year}
  {2018})\BibitemShut {NoStop}%
\bibitem [{\citenamefont {Goldenfeld}(1992)}]{Goldenfeld1992}%
  \BibitemOpen
  \bibfield  {author} {\bibinfo {author} {\bibfnamefont {N.}~\bibnamefont
  {Goldenfeld}},\ }\href {\doibase https://doi.org/10.1201/9780429493492}
  {\emph {\bibinfo {title} {{Lectures on Phase Transitions and Renormalization
  Group}}}}\ (\bibinfo  {publisher} {CRC Press},\ \bibinfo {address} {Boca
  Raton},\ \bibinfo {year} {1992})\BibitemShut {NoStop}%
\bibitem [{\citenamefont {Challet}\ \emph {et~al.}(2005)\citenamefont
  {Challet}, \citenamefont {Marsili},\ and\ \citenamefont
  {Zhang}}]{Challet2005}%
  \BibitemOpen
  \bibfield  {author} {\bibinfo {author} {\bibfnamefont {D.}~\bibnamefont
  {Challet}}, \bibinfo {author} {\bibfnamefont {M.}~\bibnamefont {Marsili}}, \
  and\ \bibinfo {author} {\bibfnamefont {Y.-C.}\ \bibnamefont {Zhang}},\
  }\href@noop {} {\emph {\bibinfo {title} {{Minority Games}}}}\ (\bibinfo
  {publisher} {Oxford University Press},\ \bibinfo {address} {New York},\
  \bibinfo {year} {2005})\BibitemShut {NoStop}%
\bibitem [{\citenamefont {Gualdi}\ \emph
  {et~al.}(2015{\natexlab{a}})\citenamefont {Gualdi}, \citenamefont {Tarzia},
  \citenamefont {Zamponi},\ and\ \citenamefont {Bouchaud}}]{Gualdi2015a}%
  \BibitemOpen
  \bibfield  {author} {\bibinfo {author} {\bibfnamefont {S.}~\bibnamefont
  {Gualdi}}, \bibinfo {author} {\bibfnamefont {M.}~\bibnamefont {Tarzia}},
  \bibinfo {author} {\bibfnamefont {F.}~\bibnamefont {Zamponi}}, \ and\
  \bibinfo {author} {\bibfnamefont {J.~P.}\ \bibnamefont {Bouchaud}},\ }\href
  {\doibase 10.1016/j.jedc.2014.08.003} {\bibfield  {journal} {\bibinfo
  {journal} {Journal of Economic Dynamics and Control}\ }\textbf {\bibinfo
  {volume} {50}},\ \bibinfo {pages} {29} (\bibinfo {year}
  {2015}{\natexlab{a}})},\ \Eprint {http://arxiv.org/abs/1307.5319}
  {arXiv:1307.5319} \BibitemShut {NoStop}%
\bibitem [{\citenamefont {Watts}\ and\ \citenamefont
  {Strogatz}(1998)}]{Watts1998}%
  \BibitemOpen
  \bibfield  {author} {\bibinfo {author} {\bibfnamefont {D.}~\bibnamefont
  {Watts}}\ and\ \bibinfo {author} {\bibfnamefont {S.}~\bibnamefont
  {Strogatz}},\ }\href {\doibase https://doi.org/10.1038/30918} {\bibfield
  {journal} {\bibinfo  {journal} {Nature}\ }\textbf {\bibinfo {volume} {393}},\
  \bibinfo {pages} {440} (\bibinfo {year} {1998})}\BibitemShut {NoStop}%
\bibitem [{\citenamefont {Gualdi}\ \emph
  {et~al.}(2015{\natexlab{b}})\citenamefont {Gualdi}, \citenamefont {Bouchaud},
  \citenamefont {Cencetti}, \citenamefont {Tarzia},\ and\ \citenamefont
  {Zamponi}}]{Gualdi2015b}%
  \BibitemOpen
  \bibfield  {author} {\bibinfo {author} {\bibfnamefont {S.}~\bibnamefont
  {Gualdi}}, \bibinfo {author} {\bibfnamefont {J.~P.}\ \bibnamefont
  {Bouchaud}}, \bibinfo {author} {\bibfnamefont {G.}~\bibnamefont {Cencetti}},
  \bibinfo {author} {\bibfnamefont {M.}~\bibnamefont {Tarzia}}, \ and\ \bibinfo
  {author} {\bibfnamefont {F.}~\bibnamefont {Zamponi}},\ }\href {\doibase
  10.1103/PhysRevLett.114.088701} {\bibfield  {journal} {\bibinfo  {journal}
  {Physical Review Letters}\ }\textbf {\bibinfo {volume} {114}},\ \bibinfo
  {pages} {088701} (\bibinfo {year} {2015}{\natexlab{b}})},\ \Eprint
  {http://arxiv.org/abs/1409.3296} {arXiv:1409.3296} \BibitemShut {NoStop}%
\bibitem [{\citenamefont {Sethna}\ \emph {et~al.}(2001)\citenamefont {Sethna},
  \citenamefont {Dahmen},\ and\ \citenamefont {Myers}}]{Sethna2001}%
  \BibitemOpen
  \bibfield  {author} {\bibinfo {author} {\bibfnamefont {J.~P.}\ \bibnamefont
  {Sethna}}, \bibinfo {author} {\bibfnamefont {K.~A.}\ \bibnamefont {Dahmen}},
  \ and\ \bibinfo {author} {\bibfnamefont {C.~R.}\ \bibnamefont {Myers}},\
  }\href {\doibase 10.1038/35065675} {\bibfield  {journal} {\bibinfo  {journal}
  {Nature}\ }\textbf {\bibinfo {volume} {410}},\ \bibinfo {pages} {242}
  (\bibinfo {year} {2001})}\BibitemShut {NoStop}%
\bibitem [{\citenamefont {Bak}(2013)}]{Bak2013}%
  \BibitemOpen
  \bibfield  {author} {\bibinfo {author} {\bibfnamefont {P.}~\bibnamefont
  {Bak}},\ }\href@noop {} {\emph {\bibinfo {title} {{How Nature Works: the
  science of self-organized criticality}}}}\ (\bibinfo  {publisher} {Springer
  Science and Business Media},\ \bibinfo {year} {2013})\BibitemShut {NoStop}%
\bibitem [{\citenamefont {Bouchaud}(2013)}]{Bouchaud2013}%
  \BibitemOpen
  \bibfield  {author} {\bibinfo {author} {\bibfnamefont {J.~P.}\ \bibnamefont
  {Bouchaud}},\ }\href {\doibase 10.1007/s10955-012-0687-3} {\bibfield
  {journal} {\bibinfo  {journal} {Journal of Statistical Physics}\ }\textbf
  {\bibinfo {volume} {151}},\ \bibinfo {pages} {567} (\bibinfo {year}
  {2013})},\ \Eprint {http://arxiv.org/abs/1209.0453} {arXiv:1209.0453}
  \BibitemShut {NoStop}%
\bibitem [{\citenamefont {Epstein}(1999)}]{Epstein1999}%
  \BibitemOpen
  \bibfield  {author} {\bibinfo {author} {\bibfnamefont {J.~M.}\ \bibnamefont
  {Epstein}},\ }\href {\doibase
  10.1002/(SICI)1099-0526(199905/06)4:5<41::AID-CPLX9>3.0.CO;2-F} {\bibfield
  {journal} {\bibinfo  {journal} {Complexity}\ }\textbf {\bibinfo {volume}
  {4}},\ \bibinfo {pages} {41} (\bibinfo {year} {1999})}\BibitemShut {NoStop}%
\bibitem [{\citenamefont {Goldstone}\ and\ \citenamefont
  {Janssen}(2005)}]{Goldstone2005}%
  \BibitemOpen
  \bibfield  {author} {\bibinfo {author} {\bibfnamefont {R.~L.}\ \bibnamefont
  {Goldstone}}\ and\ \bibinfo {author} {\bibfnamefont {M.~A.}\ \bibnamefont
  {Janssen}},\ }\href {\doibase 10.1016/j.tics.2005.07.009} {\bibfield
  {journal} {\bibinfo  {journal} {Trends in Cognitive Sciences}\ }\textbf
  {\bibinfo {volume} {9}},\ \bibinfo {pages} {424} (\bibinfo {year}
  {2005})}\BibitemShut {NoStop}%
\bibitem [{\citenamefont {Castellano}\ \emph {et~al.}(2009)\citenamefont
  {Castellano}, \citenamefont {Fortunato},\ and\ \citenamefont
  {Loreto}}]{Castellano2009}%
  \BibitemOpen
  \bibfield  {author} {\bibinfo {author} {\bibfnamefont {C.}~\bibnamefont
  {Castellano}}, \bibinfo {author} {\bibfnamefont {S.}~\bibnamefont
  {Fortunato}}, \ and\ \bibinfo {author} {\bibfnamefont {V.}~\bibnamefont
  {Loreto}},\ }\href {\doibase 10.1103/RevModPhys.81.591} {\bibfield  {journal}
  {\bibinfo  {journal} {Reviews of Modern Physics}\ }\textbf {\bibinfo {volume}
  {81}},\ \bibinfo {pages} {591} (\bibinfo {year} {2009})},\ \Eprint
  {http://arxiv.org/abs/0710.3256} {arXiv:0710.3256} \BibitemShut {NoStop}%
\bibitem [{\citenamefont {Anderson}(1988)}]{Anderson1988}%
  \BibitemOpen
  \bibfield  {author} {\bibinfo {author} {\bibfnamefont {P.~W.}\ \bibnamefont
  {Anderson}},\ }in\ \href {\doibase https://doi.org/10.1201/9780429492846}
  {\emph {\bibinfo {booktitle} {Evolutionary Paths of the Global Economy
  Workshop, Santa Fe, 1987}}}\ (\bibinfo  {publisher} {CRC Press},\ \bibinfo
  {address} {Santa Fe},\ \bibinfo {year} {1988})\BibitemShut {NoStop}%
\bibitem [{\citenamefont {Turrell}(2016)}]{Turrell2016}%
  \BibitemOpen
  \bibfield  {author} {\bibinfo {author} {\bibfnamefont {A.}~\bibnamefont
  {Turrell}},\ }\href@noop {} {\bibfield  {journal} {\bibinfo  {journal}
  {Quarterly Bulletin of the Bank of England}\ }\textbf {\bibinfo {volume}
  {Q4}},\ \bibinfo {pages} {173} (\bibinfo {year} {2016})}\BibitemShut
  {NoStop}%
\bibitem [{\citenamefont {Braun-Munzinger}\ \emph {et~al.}(2018)\citenamefont
  {Braun-Munzinger}, \citenamefont {Liu},\ and\ \citenamefont
  {Turrell}}]{Braun-Munzinger2018}%
  \BibitemOpen
  \bibfield  {author} {\bibinfo {author} {\bibfnamefont {K.}~\bibnamefont
  {Braun-Munzinger}}, \bibinfo {author} {\bibfnamefont {Z.}~\bibnamefont
  {Liu}}, \ and\ \bibinfo {author} {\bibfnamefont {A.}~\bibnamefont
  {Turrell}},\ }\href {\doibase 10.2139/ssrn.2766368} {\bibfield  {journal}
  {\bibinfo  {journal} {Bank of England Staff Working Papers}\ }\textbf
  {\bibinfo {volume} {592}} (\bibinfo {year} {2018}),\
  10.2139/ssrn.2766368}\BibitemShut {NoStop}%
\bibitem [{\citenamefont {Baptista}\ \emph {et~al.}(2016)\citenamefont
  {Baptista}, \citenamefont {Hinterschweiger}, \citenamefont {Low},\ and\
  \citenamefont {Uluc}}]{Baptista2016}%
  \BibitemOpen
  \bibfield  {author} {\bibinfo {author} {\bibfnamefont {R.}~\bibnamefont
  {Baptista}}, \bibinfo {author} {\bibfnamefont {M.}~\bibnamefont
  {Hinterschweiger}}, \bibinfo {author} {\bibfnamefont {K.}~\bibnamefont
  {Low}}, \ and\ \bibinfo {author} {\bibfnamefont {A.}~\bibnamefont {Uluc}},\
  }\href {\doibase 10.2139/ssrn.2850414} {\bibfield  {journal} {\bibinfo
  {journal} {Bank of England Staff Working Papers}\ }\textbf {\bibinfo {volume}
  {619}} (\bibinfo {year} {2016}),\ 10.2139/ssrn.2850414}\BibitemShut {NoStop}%
\bibitem [{\citenamefont {Lamperti}\ \emph {et~al.}(2018)\citenamefont
  {Lamperti}, \citenamefont {Dosi}, \citenamefont {Napoletano}, \citenamefont
  {Roventini},\ and\ \citenamefont {Sapio}}]{Lamperti2018a}%
  \BibitemOpen
  \bibfield  {author} {\bibinfo {author} {\bibfnamefont {F.}~\bibnamefont
  {Lamperti}}, \bibinfo {author} {\bibfnamefont {G.}~\bibnamefont {Dosi}},
  \bibinfo {author} {\bibfnamefont {M.}~\bibnamefont {Napoletano}}, \bibinfo
  {author} {\bibfnamefont {A.}~\bibnamefont {Roventini}}, \ and\ \bibinfo
  {author} {\bibfnamefont {A.}~\bibnamefont {Sapio}},\ }\href {\doibase
  10.2139/ssrn.3219924} {\bibfield  {journal} {\bibinfo  {journal} {Scuola
  Superiore Sant'Anna, LEM Working Papers}\ }\textbf {\bibinfo {volume} {14}}
  (\bibinfo {year} {2018}),\ 10.2139/ssrn.3219924}\BibitemShut {NoStop}%
\bibitem [{\citenamefont {Haldane}\ and\ \citenamefont
  {Turrell}(2018)}]{Haldane2018}%
  \BibitemOpen
  \bibfield  {author} {\bibinfo {author} {\bibfnamefont {A.~G.}\ \bibnamefont
  {Haldane}}\ and\ \bibinfo {author} {\bibfnamefont {A.~E.}\ \bibnamefont
  {Turrell}},\ }\href {\doibase 10.1093/oxrep/grx051} {\bibfield  {journal}
  {\bibinfo  {journal} {Oxford Review of Economic Policy}\ }\textbf {\bibinfo
  {volume} {34}},\ \bibinfo {pages} {219} (\bibinfo {year} {2018})}\BibitemShut
  {NoStop}%
\bibitem [{\citenamefont {Biere}\ \emph {et~al.}(2009)\citenamefont {Biere},
  \citenamefont {Heule},\ and\ \citenamefont {van Maaren}}]{handbook}%
  \BibitemOpen
  \bibfield  {author} {\bibinfo {author} {\bibfnamefont {A.}~\bibnamefont
  {Biere}}, \bibinfo {author} {\bibfnamefont {M.}~\bibnamefont {Heule}}, \ and\
  \bibinfo {author} {\bibfnamefont {H.}~\bibnamefont {van Maaren}},\
  }\href@noop {} {\emph {\bibinfo {title} {Handbook of satisfiability}}},\
  Vol.\ \bibinfo {volume} {185}\ (\bibinfo  {publisher} {IOS press},\ \bibinfo
  {year} {2009})\BibitemShut {NoStop}%
\bibitem [{\citenamefont {Mezard}\ and\ \citenamefont
  {Zecchina}(2002)}]{Mezard2002c}%
  \BibitemOpen
  \bibfield  {author} {\bibinfo {author} {\bibfnamefont {M.}~\bibnamefont
  {Mezard}}\ and\ \bibinfo {author} {\bibfnamefont {R.}~\bibnamefont
  {Zecchina}},\ }\href {\doibase 10.1103/PhysRevE.66.056126} {\bibfield
  {journal} {\bibinfo  {journal} {Physical Review E}\ }\textbf {\bibinfo
  {volume} {66}},\ \bibinfo {pages} {056126} (\bibinfo {year} {2002})},\
  \Eprint {http://arxiv.org/abs/0207194} {arXiv:0207194} \BibitemShut {NoStop}%
\bibitem [{\citenamefont {Altarelli}\ \emph {et~al.}(2009)\citenamefont
  {Altarelli}, \citenamefont {Monasson}, \citenamefont {Semerjian},\ and\
  \citenamefont {Zamponi}}]{Altarelli2008a}%
  \BibitemOpen
  \bibfield  {author} {\bibinfo {author} {\bibfnamefont {F.}~\bibnamefont
  {Altarelli}}, \bibinfo {author} {\bibfnamefont {R.}~\bibnamefont {Monasson}},
  \bibinfo {author} {\bibfnamefont {G.}~\bibnamefont {Semerjian}}, \ and\
  \bibinfo {author} {\bibfnamefont {F.}~\bibnamefont {Zamponi}},\ }in\ \href
  {https://arxiv.org/pdf/0802.1829.pdf} {\emph {\bibinfo {booktitle} {Handbook
  of Satisfiability}}}\ (\bibinfo  {publisher} {IOS Press, Amsterdam},\
  \bibinfo {year} {2009})\ \Eprint {http://arxiv.org/abs/0802.1829v1}
  {arXiv:0802.1829v1} \BibitemShut {NoStop}%
\bibitem [{\citenamefont {Antenucci}\ \emph {et~al.}(2019)\citenamefont
  {Antenucci}, \citenamefont {Franz}, \citenamefont {Urbani},\ and\
  \citenamefont {Zdeborov{\'{a}}}}]{Antenucci2018}%
  \BibitemOpen
  \bibfield  {author} {\bibinfo {author} {\bibfnamefont {F.}~\bibnamefont
  {Antenucci}}, \bibinfo {author} {\bibfnamefont {S.}~\bibnamefont {Franz}},
  \bibinfo {author} {\bibfnamefont {P.}~\bibnamefont {Urbani}}, \ and\ \bibinfo
  {author} {\bibfnamefont {L.}~\bibnamefont {Zdeborov{\'{a}}}},\ }\href
  {\doibase 10.1103/PhysRevX.9.011020} {\bibfield  {journal} {\bibinfo
  {journal} {Phys. Rev. X}\ }\textbf {\bibinfo {volume} {9}},\ \bibinfo {pages}
  {011020} (\bibinfo {year} {2019})},\ \Eprint
  {http://arxiv.org/abs/1805.05857} {arXiv:1805.05857} \BibitemShut {NoStop}%
\bibitem [{\citenamefont {Franz}\ \emph {et~al.}(2017)\citenamefont {Franz},
  \citenamefont {Parisi}, \citenamefont {Sevelev}, \citenamefont {Urbani},\
  and\ \citenamefont {Zamponi}}]{Franz2017a}%
  \BibitemOpen
  \bibfield  {author} {\bibinfo {author} {\bibfnamefont {S.}~\bibnamefont
  {Franz}}, \bibinfo {author} {\bibfnamefont {G.}~\bibnamefont {Parisi}},
  \bibinfo {author} {\bibfnamefont {M.}~\bibnamefont {Sevelev}}, \bibinfo
  {author} {\bibfnamefont {P.}~\bibnamefont {Urbani}}, \ and\ \bibinfo {author}
  {\bibfnamefont {F.}~\bibnamefont {Zamponi}},\ }\href {\doibase
  10.21468/SciPostPhys.2.3.019} {\bibfield  {journal} {\bibinfo  {journal}
  {SciPost Phys.}\ }\textbf {\bibinfo {volume} {019}},\ \bibinfo {pages} {1}
  (\bibinfo {year} {2017})}\BibitemShut {NoStop}%
\bibitem [{\citenamefont {Sharma}\ \emph {et~al.}(2019)\citenamefont {Sharma},
  \citenamefont {Bouchaud}, \citenamefont {Tarzia},\ and\ \citenamefont
  {Zamponi}}]{Sharma2019}%
  \BibitemOpen
  \bibfield  {author} {\bibinfo {author} {\bibfnamefont {D.}~\bibnamefont
  {Sharma}}, \bibinfo {author} {\bibfnamefont {J.~P.}\ \bibnamefont
  {Bouchaud}}, \bibinfo {author} {\bibfnamefont {M.}~\bibnamefont {Tarzia}}, \
  and\ \bibinfo {author} {\bibfnamefont {F.}~\bibnamefont {Zamponi}},\ }\href
  {\doibase https://doi.org/10.1088/1742-5468/ab4800} {\bibfield  {journal}
  {\bibinfo  {journal} {Journal of Statistical Mechanics: Theory and
  Experiment}\ ,\ \bibinfo {pages} {123301}} (\bibinfo {year} {2019})},\
  \Eprint {http://arxiv.org/abs/1906.01490} {arXiv:1906.01490} \BibitemShut
  {NoStop}%
\bibitem [{\citenamefont {Gaffeo}\ \emph {et~al.}(2008)\citenamefont {Gaffeo},
  \citenamefont {{Delli Gatti}}, \citenamefont {Desiderio},\ and\ \citenamefont
  {Gallegati}}]{Gaffeo2008}%
  \BibitemOpen
  \bibfield  {author} {\bibinfo {author} {\bibfnamefont {E.}~\bibnamefont
  {Gaffeo}}, \bibinfo {author} {\bibfnamefont {D.}~\bibnamefont {{Delli
  Gatti}}}, \bibinfo {author} {\bibfnamefont {S.}~\bibnamefont {Desiderio}}, \
  and\ \bibinfo {author} {\bibfnamefont {M.}~\bibnamefont {Gallegati}},\ }\href
  {\doibase 10.1057/eej.2008.27} {\bibfield  {journal} {\bibinfo  {journal}
  {Eastern Economic Journal}\ }\textbf {\bibinfo {volume} {34}},\ \bibinfo
  {pages} {441} (\bibinfo {year} {2008})}\BibitemShut {NoStop}%
\bibitem [{\citenamefont {Rosenblatt}(1958)}]{Rosenblatt1958}%
  \BibitemOpen
  \bibfield  {author} {\bibinfo {author} {\bibfnamefont {F.}~\bibnamefont
  {Rosenblatt}},\ }\href@noop {} {\bibfield  {journal} {\bibinfo  {journal}
  {Psychological Review}\ }\textbf {\bibinfo {volume} {65}},\ \bibinfo {pages}
  {386} (\bibinfo {year} {1958})}\BibitemShut {NoStop}%
\bibitem [{\citenamefont {Gardner}(1988)}]{Gardner1988}%
  \BibitemOpen
  \bibfield  {author} {\bibinfo {author} {\bibfnamefont {E.}~\bibnamefont
  {Gardner}},\ }\href {\doibase 10.1088/0305-4470/21/1/030} {\bibfield
  {journal} {\bibinfo  {journal} {Journal of Physics A: Mathematical and
  General}\ }\textbf {\bibinfo {volume} {21}},\ \bibinfo {pages} {257}
  (\bibinfo {year} {1988})}\BibitemShut {NoStop}%
\bibitem [{\citenamefont {Gardner}\ and\ \citenamefont
  {Derrida}(1988)}]{Gardner1988a}%
  \BibitemOpen
  \bibfield  {author} {\bibinfo {author} {\bibfnamefont {E.}~\bibnamefont
  {Gardner}}\ and\ \bibinfo {author} {\bibfnamefont {B.}~\bibnamefont
  {Derrida}},\ }\href {\doibase 10.1088/0305-4470/21/1/031} {\bibfield
  {journal} {\bibinfo  {journal} {Journal of Physics A: Mathematical and
  General}\ }\textbf {\bibinfo {volume} {21}},\ \bibinfo {pages} {271}
  (\bibinfo {year} {1988})}\BibitemShut {NoStop}%
\bibitem [{\citenamefont {Krauth}\ \emph {et~al.}(1988)\citenamefont {Krauth},
  \citenamefont {Mezard},\ and\ \citenamefont {Nadal}}]{Krauth1988}%
  \BibitemOpen
  \bibfield  {author} {\bibinfo {author} {\bibfnamefont {W.}~\bibnamefont
  {Krauth}}, \bibinfo {author} {\bibfnamefont {M.}~\bibnamefont {Mezard}}, \
  and\ \bibinfo {author} {\bibfnamefont {J.~P.}\ \bibnamefont {Nadal}},\ }\href
  {http://dl.acm.org/citation.cfm?id=56123.56124} {\bibfield  {journal}
  {\bibinfo  {journal} {Complex Systems}\ }\textbf {\bibinfo {volume} {2}},\
  \bibinfo {pages} {387} (\bibinfo {year} {1988})}\BibitemShut {NoStop}%
\bibitem [{\citenamefont {Krauth}\ and\ \citenamefont
  {M{\'{e}}zard}(1989)}]{Krauth1989}%
  \BibitemOpen
  \bibfield  {author} {\bibinfo {author} {\bibfnamefont {W.}~\bibnamefont
  {Krauth}}\ and\ \bibinfo {author} {\bibfnamefont {M.}~\bibnamefont
  {M{\'{e}}zard}},\ }\href {\doibase 10.1051/jphys:0198900500200305700}
  {\bibfield  {journal} {\bibinfo  {journal} {Journal de Physique}\ }\textbf
  {\bibinfo {volume} {50}},\ \bibinfo {pages} {3057} (\bibinfo {year}
  {1989})}\BibitemShut {NoStop}%
\bibitem [{\citenamefont {Brunel}\ \emph {et~al.}(1992)\citenamefont {Brunel},
  \citenamefont {Nadal},\ and\ \citenamefont {Toulouse}}]{Brunel1992}%
  \BibitemOpen
  \bibfield  {author} {\bibinfo {author} {\bibfnamefont {N.}~\bibnamefont
  {Brunel}}, \bibinfo {author} {\bibfnamefont {J.~P.}\ \bibnamefont {Nadal}}, \
  and\ \bibinfo {author} {\bibfnamefont {G.}~\bibnamefont {Toulouse}},\ }\href
  {\doibase https://doi.org/10.1088/0305-4470/25/19/015} {\bibfield  {journal}
  {\bibinfo  {journal} {J. Phys. A}\ }\textbf {\bibinfo {volume} {25}},\
  \bibinfo {pages} {5017} (\bibinfo {year} {1992})}\BibitemShut {NoStop}%
\bibitem [{\citenamefont {Marsili}(2014)}]{Marsili2014}%
  \BibitemOpen
  \bibfield  {author} {\bibinfo {author} {\bibfnamefont {M.}~\bibnamefont
  {Marsili}},\ }\href {\doibase 10.1080/14697688.2013.765061} {\bibfield
  {journal} {\bibinfo  {journal} {Quantitative Finance}\ }\textbf {\bibinfo
  {volume} {14}},\ \bibinfo {pages} {1663} (\bibinfo {year}
  {2014})}\BibitemShut {NoStop}%
\bibitem [{Note1()}]{Note1}%
  \BibitemOpen
  \bibinfo {note} {Our simulations were in fact done with the lower bound on
  prices i.e. $x_{m}=0.01$ and not zero. This has no material impact on the
  present results however.}\BibitemShut {Stop}%
\bibitem [{\citenamefont {Yanagita}\ and\ \citenamefont
  {Onozaki}(2010)}]{Yanagita2010}%
  \BibitemOpen
  \bibfield  {author} {\bibinfo {author} {\bibfnamefont {T.}~\bibnamefont
  {Yanagita}}\ and\ \bibinfo {author} {\bibfnamefont {T.}~\bibnamefont
  {Onozaki}},\ }\href {\doibase 10.1016/j.physa.2009.10.040} {\bibfield
  {journal} {\bibinfo  {journal} {Physica A: Statistical Mechanics and its
  Applications}\ }\textbf {\bibinfo {volume} {389}},\ \bibinfo {pages} {1041}
  (\bibinfo {year} {2010})}\BibitemShut {NoStop}%
\bibitem [{\citenamefont {Galluccio}\ \emph {et~al.}(1998)\citenamefont
  {Galluccio}, \citenamefont {Bouchaud},\ and\ \citenamefont
  {Potters}}]{galluccio1998rational}%
  \BibitemOpen
  \bibfield  {author} {\bibinfo {author} {\bibfnamefont {S.}~\bibnamefont
  {Galluccio}}, \bibinfo {author} {\bibfnamefont {J.-P.}\ \bibnamefont
  {Bouchaud}}, \ and\ \bibinfo {author} {\bibfnamefont {M.}~\bibnamefont
  {Potters}},\ }\href@noop {} {\bibfield  {journal} {\bibinfo  {journal}
  {Physica A: Statistical Mechanics and its Applications}\ }\textbf {\bibinfo
  {volume} {259}},\ \bibinfo {pages} {449} (\bibinfo {year}
  {1998})}\BibitemShut {NoStop}%
\bibitem [{\citenamefont {Ledger}\ and\ \citenamefont
  {Sojmark}(2018)}]{Ledger2018}%
  \BibitemOpen
  \bibfield  {author} {\bibinfo {author} {\bibfnamefont {S.}~\bibnamefont
  {Ledger}}\ and\ \bibinfo {author} {\bibfnamefont {A.}~\bibnamefont
  {Sojmark}},\ }\href {http://arxiv.org/abs/1807.05126} {\enquote {\bibinfo
  {title} {{At the Mercy of the Common Noise: Blow-ups in a Conditional
  McKean--Vlasov Problem}},}\ } (\bibinfo {year} {2018}),\ \Eprint
  {http://arxiv.org/abs/1807.05126} {arXiv:1807.05126} \BibitemShut {NoStop}%
\bibitem [{\citenamefont {Pangallo}(2020)}]{Pangallo2020}%
  \BibitemOpen
  \bibfield  {author} {\bibinfo {author} {\bibfnamefont {M.}~\bibnamefont
  {Pangallo}},\ }\href@noop {} {\enquote {\bibinfo {title} {{Synchronization of
  Endogenous Business Cycles}},}\ } (\bibinfo {year} {2020}),\ \Eprint
  {http://arxiv.org/abs/2002.06555} {arXiv:2002.06555} \BibitemShut {NoStop}%
\bibitem [{\citenamefont {Jorda}\ \emph {et~al.}(2011)\citenamefont {Jorda},
  \citenamefont {Schularick},\ and\ \citenamefont {Taylor}}]{Jorda2011}%
  \BibitemOpen
  \bibfield  {author} {\bibinfo {author} {\bibfnamefont {O.}~\bibnamefont
  {Jorda}}, \bibinfo {author} {\bibfnamefont {M.}~\bibnamefont {Schularick}}, \
  and\ \bibinfo {author} {\bibfnamefont {A.~M.}\ \bibnamefont {Taylor}},\
  }\href {\doibase 10.3386/w17621} {\bibfield  {journal} {\bibinfo  {journal}
  {NBER Working Paper Series}\ }\textbf {\bibinfo {volume} {17621}} (\bibinfo
  {year} {2011}),\ 10.3386/w17621}\BibitemShut {NoStop}%
\bibitem [{\citenamefont {Adrian}\ and\ \citenamefont
  {Boyarchenko}(2012)}]{Adrian2012}%
  \BibitemOpen
  \bibfield  {author} {\bibinfo {author} {\bibfnamefont {T.}~\bibnamefont
  {Adrian}}\ and\ \bibinfo {author} {\bibfnamefont {N.}~\bibnamefont
  {Boyarchenko}},\ }\href {\doibase 10.2139/ssrn.2133385} {\bibfield  {journal}
  {\bibinfo  {journal} {Becker Friedman Institute for Research in Economics
  Working Paper}\ }\textbf {\bibinfo {volume} {010}} (\bibinfo {year} {2012}),\
  10.2139/ssrn.2133385}\BibitemShut {NoStop}%
\bibitem [{\citenamefont {Aymanns}\ and\ \citenamefont
  {Farmer}(2015)}]{Aymanns2015}%
  \BibitemOpen
  \bibfield  {author} {\bibinfo {author} {\bibfnamefont {C.}~\bibnamefont
  {Aymanns}}\ and\ \bibinfo {author} {\bibfnamefont {J.~D.}\ \bibnamefont
  {Farmer}},\ }\href {\doibase 10.1016/j.jedc.2014.09.015} {\bibfield
  {journal} {\bibinfo  {journal} {Journal of Economic Dynamics and Control}\
  }\textbf {\bibinfo {volume} {50}},\ \bibinfo {pages} {155} (\bibinfo {year}
  {2015})},\ \Eprint {http://arxiv.org/abs/1407.5305} {arXiv:1407.5305}
  \BibitemShut {NoStop}%
\bibitem [{\citenamefont {Aymanns}\ \emph {et~al.}(2016)\citenamefont
  {Aymanns}, \citenamefont {Caccioli}, \citenamefont {Farmer},\ and\
  \citenamefont {Tan}}]{Aymanns2016}%
  \BibitemOpen
  \bibfield  {author} {\bibinfo {author} {\bibfnamefont {C.}~\bibnamefont
  {Aymanns}}, \bibinfo {author} {\bibfnamefont {F.}~\bibnamefont {Caccioli}},
  \bibinfo {author} {\bibfnamefont {J.~D.}\ \bibnamefont {Farmer}}, \ and\
  \bibinfo {author} {\bibfnamefont {V.~W.}\ \bibnamefont {Tan}},\ }\href
  {\doibase 10.1016/j.jfs.2016.02.004} {\bibfield  {journal} {\bibinfo
  {journal} {Journal of Financial Stability}\ }\textbf {\bibinfo {volume}
  {27}},\ \bibinfo {pages} {263} (\bibinfo {year} {2016})},\ \Eprint
  {http://arxiv.org/abs/1507.04136} {arXiv:1507.04136} \BibitemShut {NoStop}%
\bibitem [{\citenamefont {Seppecher}\ and\ \citenamefont
  {Salle}(2015)}]{Seppecher2015}%
  \BibitemOpen
  \bibfield  {author} {\bibinfo {author} {\bibfnamefont {P.}~\bibnamefont
  {Seppecher}}\ and\ \bibinfo {author} {\bibfnamefont {I.}~\bibnamefont
  {Salle}},\ }\href {\doibase 10.1080/00036846.2015.1021456} {\bibfield
  {journal} {\bibinfo  {journal} {Applied Economics}\ }\textbf {\bibinfo
  {volume} {47}},\ \bibinfo {pages} {3771} (\bibinfo {year}
  {2015})}\BibitemShut {NoStop}%
\end{thebibliography}%

\end{document}